\documentclass[a4paper,fleqn]{cas-dc}
\usepackage[numbers]{natbib}
\usepackage{amsmath,amssymb}
\usepackage[cp1251]{inputenc}

\usepackage{srctex}
\usepackage{hyperref}

\begin{document}
\let\WriteBookmarks\relax
\def\floatpagepagefraction{1}
\def\textpagefraction{.001}
\shorttitle{Deep machine learning potentials for multicomponent metallic melts}
\shortauthors{R.E.~Ryltsev et~al.}


\title [mode = title]{Deep machine learning potentials for multicomponent metallic melts: development, predictability and compositional transferability}
\tnotemark[1]


\author[1,2,3]{R.E.~Ryltsev}[orcid=0000-0003-1746-8200]
\cormark[1]
\ead{rrylcev@mail.ru}

\author[3]{N.M.~Chtchelkatchev}[orcid=0000-0002-7242-1483]
\ead[url]{chtchelkatchev@hppi.troitsk.ru}

\address [1]{Institute of Metallurgy of the Ural Branch of the Russian Academy of Sciences, 620016, Ekaterinburg, Russia}
\address [2]{Ural Federal University, 620002, Ekaterinburg, Russia}
\address [2]{Vereshchagin Institute for High Pressure Physics, Russian Academy of Sciences, 108840 Troitsk, Moscow, Russia}


\begin{abstract}
The use of machine learning interatomic potentials (MLIPs) in simulations of materials is a state-of-the-art approach, which allows achieving nearly \textit{ab initio} accuracy with orders of magnitude less computational cost. Multicomponent disordered systems have a highly complicated potential energy surface due to both topological and compositional disorder. That arises issues in MLIPs developing, such as optimal design strategy of potentials and their predictability and transferability. Here we address MLIPs for multicomponent metallic melts taking the ternary Al-Cu-Ni ones as a convenient example. We use many-body deep machine learning potentials as implemented in the DeePMD-kit to build MLIP that allows describing both atomic structure and dynamics of the system in the whole composition range. Doing that we consider different sets of neural networks hyperparameters and learning schemes to create an optimal MLIP, which allows archiving good accuracy in comparison with both \textit{ab initio} and experimental data. We find that developed MLIP demonstrates good compositional transferability, which extends far beyond compositional fluctuations in the training configurations. The results obtained open up prospects for simulating structural and dynamical properties of multicomponent metallic alloys with MLIPs.
\end{abstract}


\begin{highlights}

\item We develop deep machine learning potential for Al-Cu-Ni melts in the whole composition range
\item The potential provides excellent accuracy in comparison to ab initio and experimental data
\item The potential reveals good compositional transferability
\end{highlights}


\maketitle

\section{Introduction}

The use of machine learning is a new paradigm in modern computational materials science~\cite{Maleki2021JMolLiq,Schmidt2019,Morgan2020AnnRevMaterRes,Sepehri2020JMolLiq,Schleder2019JPhysMater,Wei2019InfoMat,Chng2017PRX}. One of the most promising and widely accepted techniques is using \textit{ab initio} reference data on energies, forces, and stress tensors to develop machine learning interatomic potentials (MLIPs) with a flexible functional form  which can effectively fit the potential energy surface (PES) of the particle system~\cite{Ceriotti2021JCP,Mishin2021ActaMater,VonLilienfeld2020NatComm,Behler2021EPJ,Deringer2020JPhysEnerg,Mueller2020JCP,Deringer2019AdvMater,Behler2016JCP}. This approach allows solving the principle problems of \textit{ab initio} simulations: the effects of ``small size'' and ``short time'', associated with difficulties in simulating sufficiently large supercells at long enough computational times. MLIPs can provide nearly \textit{ab initio} accuracy with orders of magnitude less computational cost for systems composed of up to millions of particles~\cite{Lu2021CompPhysComm}.

One of the most challenging applications of MLIPs is molecular dynamics simulations of disordered systems, such as melts, supercooled liquids, and glasses. Developing accurate MLIPs for such systems is a more difficult task than for crystals. Indeed, due to the absence of long-range order, it is necessary to consider rather large supercells that makes \textit{ab initio} simulations very costly or even hardly possible. Moreover, the number of possible structural configuration in disordered systems are enormously large and so it is more difficult to sample the configurational space and build representative training dataset.

Despite these difficulties, a lot of effective  MLIPs, which describe perfectly the disordered systems of different nature, have been recently developed \cite{Deringer2021Nature,Deringer2020NatComm,Gartner2020PNAS,Balyakin2020PRE,Sommers2020PCCP,Niu2020NatComm,Wen2019PRB,Tang2020PCCP,Zhou2021PhysStatSol}. However, these examples are either unary or binary systems; multicomponent systems are studied relatively weak.

Multicomponent disordered systems have a highly complicated PES due to both topological and compositional disorder. Therefore, a model describing interaction in such systems has to contain a large set of parameters especially when we pretend to describe the whole composition range. It is difficult to build such universal interactions using simple models like EAM ones; more flexible (so-called mathematical) potentials are needed here. There are three major classes of regressors, which are used to build MLIPs: deep neural networks \cite{Zhang2019PRM,Wen2019PRB,Zhang2018PRL,Zhang2020CompPhysComm,Singraber2019JChemTheorComp,Behler2016JCP,Schut2017NatureComm,Smith2017ChemSci,Gao2020JCnemInfMod,Shrutt2018JCP}, kernel methods ~\cite{Bartok2010PRL,Bartok2015IJQC,Jinnouchi2019PRL,Jinnouchi2019PRL-2} and generalized linear models~\cite{Thompson2015JCompPhys,Li2018PRB,Novoselov2019CompMaterSci,Novikov2021MLSciTech}. There are also promising alternative approaches, which are not widely accepted so far \cite{Mueller2020JCP,Mishin2021ActaMater}. We argue that deep neural network potentials (DNNPs) are the most suitable models for this challenging problem. These neural networks have high flexibility and so can be treated as universal approximators~\cite{Hornil1989NN}. Among different DNNP-based approaches, the DeePMD-kit is one of the most convenient and powerful ones~\cite{Wang2018CompPhysComm,Zhang2018DPMDSE}. The main advantages of DeePMD are: 1) highly effective and automatic procedure to map the particles coordinates to the space of structural descriptors, which includes many-body interactions and provides invariance with respect to translations, rotations, and permutations; 2) the use of the TensorFlow platform to build models allows using GPU to effectively train DNNPs and than utilize them in large-scale molecular dynamics simulations~\cite{Lu2021CompPhysComm}; 3) the possibility to apply active learning strategy using the DPGEN tool~\cite{Zhang2020CompPhysComm,Zhang2019PRM}. This approach has been successfully utilized for simulating systems of different nature~\cite{Niu2020NatComm,Sommers2020PCCP,Gartner2020PNAS,Balyakin2020PRE,Wen2019PRB,Tang2020PCCP,Zhang2021PRL,Andolina2020JCP}. However, there are no implicit receipts how to choose values of DNNP hyperparameters (such as the number of hidden layers and the number of neurons) as well as the learning parameters (such as learning rates, batch size, etc.) that are optimal to simulate multicomponent melts. A vast number of different parameter combinations exist, which can produce models differing substantially by their accuracy and computational efficiency (see Tab.~\ref{tab:model_parameters} below).

The key points in developing any MLIPs are their predictability and transferability. Predictability is an accuracy of a MLIP in describing different observable properties. Transferability stands for the possibility to use a model at values of thermodynamic parameters that are noticeably distinct from those included in a training dataset. In the case of multicomponent alloys, an important particular issue is \textit{compositional transferability} that is the possibility to deviate from the compositions corresponding to training configurations. The related issue is how dense should be a concentration grid of training configurations to build a MLIP, which provides an accurate description of a multicomponent alloy in the whole compositional range.

Here we address the above issues for ternary Al-Cu-Ni melts, which are convenient model systems due to several reasons. First, they are mixtures of metals with different electronic configurations: $p$-metal (Al), noble $s$-metal and transition 3$d$-metal (Ni). Therefore, the system demonstrates complex chemical interaction between components that causes non-trivial behavior, for example, the existence of complex intermetallic phases~\cite{Raghavan2006JPhaseEqDiff} and non-monotonous behavior of physicochemical properties of the melts with the change of the composition~\cite{Kamaeva2020JPCM}. Second, limiting cases of the system, such as binary Al-Ni, Al-Cu alloys, and pure Al, Cu, Ni are of great fundamental interest and practical importance~\cite{Rajkumar2020JMolLiq,Kirova2020CompMatSci,Fleita2020JPCM,Dubinin2021Metals,Trybula2018JMaterSci,Trybula2018JMaterSci,Khusnutdinoff2016JETP}.

\section{Methods}

\subsection{\textit{Ab initio} simulations and training dataset}

The training dataset for developing DNNPs is generated by \textit{ab initio} molecular dynamics (AIMD) simulations utilizing density functional theory (DFT) as implemented in Vienna \textit{ab initio} simulation program (VASP)~\cite{VASP1}. Projector augmented-wave (PAW) pseudopotentials and Perdew-Burke-Ernzerhof (PBE)~\cite{Perdew1992PRB1,Perdew1992PRB2} gradient approximation to the exchange-correlation functional~\cite{Kresse1999PRB} are applied. We use rather large supercells of $N=512$ particles. Only the $\Gamma$ point is used to sample the Brillouin zone and the energy cutoff of 300 eV is set. The AIMD simulations are performed using the canonical ensemble (NVT) with Nos\'{e}-Hoover thermostat under periodic boundary conditions. At each composition studied, we impose an equilibrium density and a temperature, which is slightly above the corresponding melting (liquidus) point. The protocol to prepare equilibrium initial configurations is described in Ref.~\cite{Kamaeva2020JMolLiq}.

\begin{figure}[pos=t]
  \centering
  \includegraphics[width=1\columnwidth]{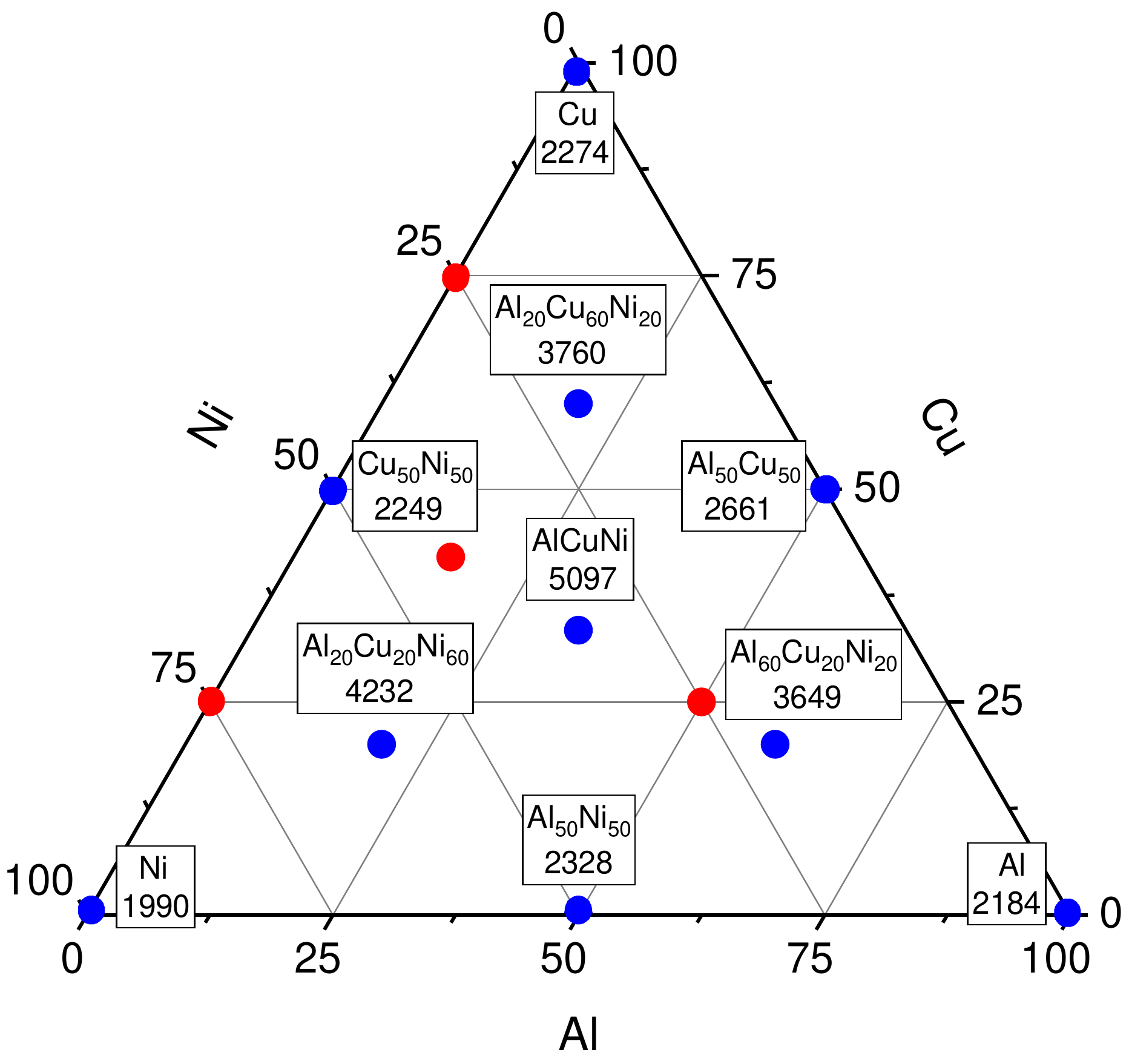}
  \caption{Ternary plot for Al-Cu-Ni system with the compositions included in the training dataset (blue bullets) and the corresponding numbers of AIMD configurations for each composition. Testing compositions, which are used to analyze compositional transferability of the DDNP, are presented by red bullets.}
  \label{fig:ternary_plot}
\end{figure}

To build the training dataset, we simulate ten Al-Cu-Ni alloys whose concentrations are evenly distributed on the whole range of compositions. The following alloys have been included: pure Al, Cu, Ni, binary ${\rm Al_{50}Cu_{50}}$, ${\rm Al_{50}Ni_{50}}$, ${\rm Cu_{50}Ni_{50}}$, and ternary ${\rm Al_{60}Cu_{20}Ni_{20}}$, ${\rm Al_{20}Cu_{60}Ni_{20}}$, \\${\rm Al_{20}Cu_{20}Ni_{60}}$, ${\rm Al_{33.3}Cu_{33.3}Ni_{33.3}}$. The latter system is an equiatomic middle-entropy alloy whose study is of special interest in the context of the new paradigm of multi-principal alloys~\cite{George2019NatMaterRev}. To illustrate better the structure of the training dataset, we draw the included compositions on the ternary plot together with the number of training AIMD configurations for each point (Fig.~\ref{fig:ternary_plot}). The whole dataset includes $N_c= 30,424$ AIMD configurations and corresponding values of total energy, interatomic forces, and virials. This dataset was split into training and testing ones in the ratio 3:1.


\subsection{Training procedure}
\begin{table*}
\caption{Details of networks structure, training scheme as well as accuracy and computational efficiency of resulted DNNPs. The vectors determining the structure of both embedding and fitting nets contain the numbers of the neurons in each network layer. The cutoffs vector is $(r_{\rm cs}, r_{\rm c})$, where $r_{\rm cs}$ is the smooth cutoff parameter and $r_{\rm c}$ is the cutoff radius of the model.   Accuracy is reflected by the vector of RMSEs for energies, forces, and virials calculated by DNNP in comparison with those from AIMD simulations. The units of RMSE are meV/atom for energies and virials and  meV/{\AA} for forces. The computational performance is expressed in the ns/day for test simulation of $N=4096$ particle AlCuNi alloy performed on single GPU.}\label{tab:model_parameters}.
\centering
\begin{tabular}{ l | c  c | c  c  | c || c  c  }
\hline
\hline
\# & embedding  & axis & fitting  & cutoffs & learning & accuracy  & perfomance  \\
   &      net   & neuron&   net   &  ({\AA})& scheme  &  (RMSE)    &  (ns/day)    \\
\hline

\hline

    1  &  (50, 100)      & 8 &    (240, 240, 240)      &(5,6)& (0.2, 0.2, 500, 500,0.0,0.0)   & (1.35, 71.1, 150.9)   & 0.37 \\
    2  &  (50, 100)      & 8 &    (240, 240, 240)      &(5,6)& (0.2, 0.2, 500, 500,0.1,0.1)   & (1.74, 68.1, 6.22)    & 0.37 \\
    3  &  (25, 50, 100)  & 8 &    (240, 240, 240)      &(5,6)& (0.2, 0.2, 500, 500,0.1,0.1)   & (1.6,  70.7, 6.32)    & 0.32\\
    4  &  (50, 100)      & 8 &    (50, 50)             &(5,6)& (0.2, 0.2, 500, 500,0.1,0.1)   & (1.65, 72.7, 6.28)    & 0.39 \\
    5  &  (50, 100)      & 8 &    (240, 240, 240, 240) &(5,6)& (0.2, 0.2, 500, 500,0.1,0.1)   & (1.7,  71.9, 6.24)    & 0.36\\
    6  &  (10, 20)       & 8 &    (240, 240, 240)      &(5,6)& (0.2, 0.2, 500, 500,0.1,0.1)   & (2.67, 115.1, 14.6)   & 1.08 \\
    7  &  (30, 60)       & 8 &    (120, 120, 120)      &(5,6)& (0.2, 0.2, 500, 500,0.1,0.1)   & (2.1,  75.2, 7.0)     & 0.67 \\
    8  &  (10, 20)       & 4 &    (50, 50)             &(5,6)& (0.2, 0.2, 500, 500,0.1,0.1)   & (3.24, 113.4, 15.79)  & 1.2\\
    9  &  (50, 100)      & 6 &    (240, 240, 240)      &(5,6)& (0.2, 0.2, 500, 500,0.1,0.1)   & (1.68, 71.1, 6.27)    & 0.37 \\
    10  &  (50, 100)      & 10 &    (240, 240, 240)    &(5,6)& (0.2, 0.2, 500, 500,0.1,0.1)   & (1.85, 67.8, 6.19)    & 0.36 \\
    11  &  (50, 100)      & 8 &    (240, 240, 240)     &(5.8,6)& (0.2, 0.2, 500, 500,0.1,0.1) & (1.73, 70.9, 6.6)     & 0.37 \\
    12  &  (50, 100)      & 8 &    (240, 240, 240)     &(6,7)& (0.2, 0.2, 500, 500,0.1,0.1)   & (1.59, 69.7, 6.51)    & 0.27 \\
    13  &  (60, 120)      & 8 &    (240, 240, 240)     &(6,7)& (0.2, 0.2, 500, 500,0.1,0.1)   & (1.51, 67.6, 5.83)    & 0.24 \\
    14  &  (40, 80, 160)   & 8 &    (240, 240, 240)     &(7,8)& (0.2, 0.2, 500, 500,0.1,0.1)   & (1.51, 69.9, 6.25)    & 0.007 \\

\hline
\hline
\end{tabular}
\end{table*}

To develop DNNPs for Al-Cu-Ni melts, we use Deep Potential Molecular Dynamic package (DeePMD-kit)~\cite{Zhang2018PRL,Wang2018CompPhysComm,Zhang2018DPMDSE,Zhang2019PRM,Wen2019PRB}. The approach utilizes feedforward multilayer neural networks as a regression model. The key problem in any MLIP developing method is to transform atomic coordinates to a set of structural descriptors, which preserve translational, rotational, and permutation symmetries. We used the smooth version (DeepPot-SE)~\cite{Zhang2018DPMDSE} of the DeePMD-kit where this transformation is performed by employing an end-to-end smooth and continuous embedding network. The loss function of the model takes into account energies, forces and virials those weights at the start and the end of the training procedure are respectively determined by the parameters $p_e^{\rm start}$, $p_e^{\rm limit}$, $p_f^{\rm start}$, $p_f^{\rm limit}$, $p_v^{\rm start}$, $p_v^{\rm limit}$. Besides, there are the decay rate $\lambda$ and decay step $n_d$, which determine how the learning rate (that is the prefactor of the gradient of the loss function) changes during the training. Other important parameters are the cutoff radius of the model $r_c$ and the smooth cutoff parameter $r_{\rm cs}$, which determines the decay distance of the descriptors. Thus, there are many hyperparameters of the embedding and the fitting networks as well as parameters of the training procedure, which have to be tuned. The total number of possible parameters combination is enormously large and it is hardly possible to search through all of them in every particular case. However, some general recommendations for the choice of main parameters should be given for a certain class of the system. Here we make a step on this way for multicomponent metallic alloys trying to tune the most important hyperparameters of the neural networks and some parameters of the learning procedure. Doing that, we train several models using the same dataset described above. The parameters of the models are listed in the Tab.~\ref{tab:model_parameters} (see detailed discussion in section~\ref{sec:param}.

\begin{figure}[pos=t]
  \centering
  \includegraphics[width=0.8\columnwidth]{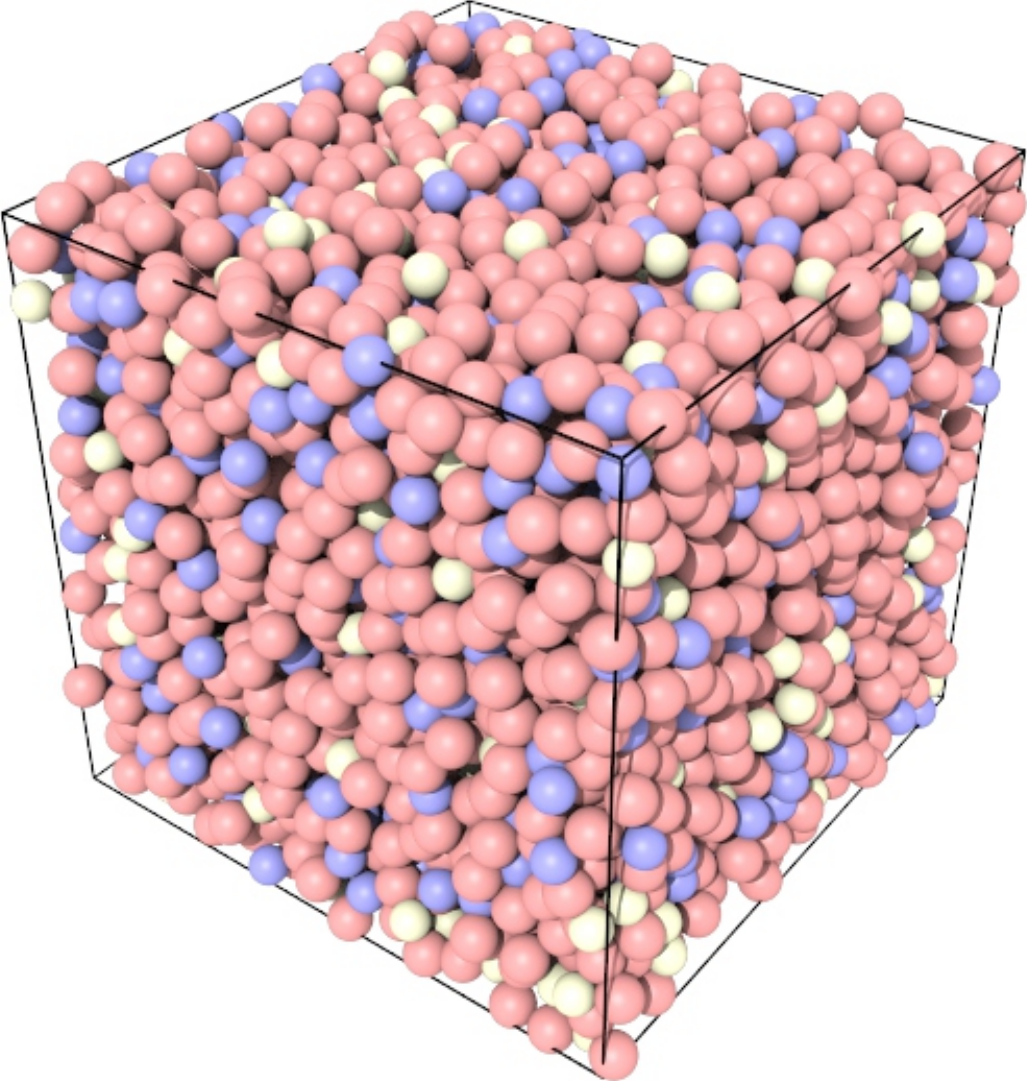}
  \caption{Snapshot of Al$_{0.7}$Cu$_{0.2}$Ni$_{0.1}$ melt with 4096 particles in the cell (Al -- red, Cu -- blue, and Ni -- light green) obtained using DNNP-based simulations.}
  \label{fig:AlCuNi}
\end{figure}
\subsection{NNP simulations}
The DeePMD-kit includes implementation to LAMMPS code~\cite{Plimpton1995JCompPhys}, which allows performing direct MD simulations with developed DNNPs. When comparing the results obtained by DNNP and AIMD calculations, to exclude the impact of finite-size effects and difference in the statistics, we apply the same simulation conditions in both cases (number of particles, number of MD steps, etc.). When calculating observable properties to compare with the experimental data, we perform simulations at much larger space-time scales: $N=4096$ particles with about 50,000 MD steps of 1 fs. In the case of NVT/NPT simulations, the Nose-Hoover thermostat/barostat with damping parameter 100/1000 fs was used to control the temperature/pressure.

The preparation of initial configurations for MLIP-based simulations deserves special discussion. When using empirical force fields like EAM potentials, due to internal physical restrictions, initial configurations can be almost arbitrary. Even where initial configurations are nonphysical (for example, strong atom overlapping takes place), their subsequent relaxation usually leads the system to a state appropriate to start MD simulations. Unfortunately, this is not the case for MLIPs, particularly DNNPs. Indeed, if we put the system into a state that differs essentially from training configurations, then DNPP has to extrapolate the values of energy and forces. Substantial extrapolation can lead to critical numerical artifacts like an irreversible agglomeration of particles at unphysically small distances. To avoid such situations, we prepare initial states from AIMD configurations included in the training dataset. We use either the AIMD configurations themselves or their replications. For example, 4096-particle systems, which was simulated to compare DNNP results with experimental data (see Figs.~\ref{fig:RDF_exp},\ref{fig:dens}), were initially prepared as $2\times2\times2$ replication of 512-particle AIMD supercell. When studying compositional transferability (see Sec.\ref{sec:comp_trans}), we simulate alloys that are not included in the training dataset. In this case, we use the same protocol as for the alloys included in the dataset. Careful analysis of the configurations does not reveal any artifacts (see an example of snapshot in Fig.~\ref{fig:AlCuNi}). That gives further evidence for a good compositional transferability of DNNPs.

\section{Results and discussion}

\begin{figure}
  \centering
  \includegraphics[width=0.8\columnwidth]{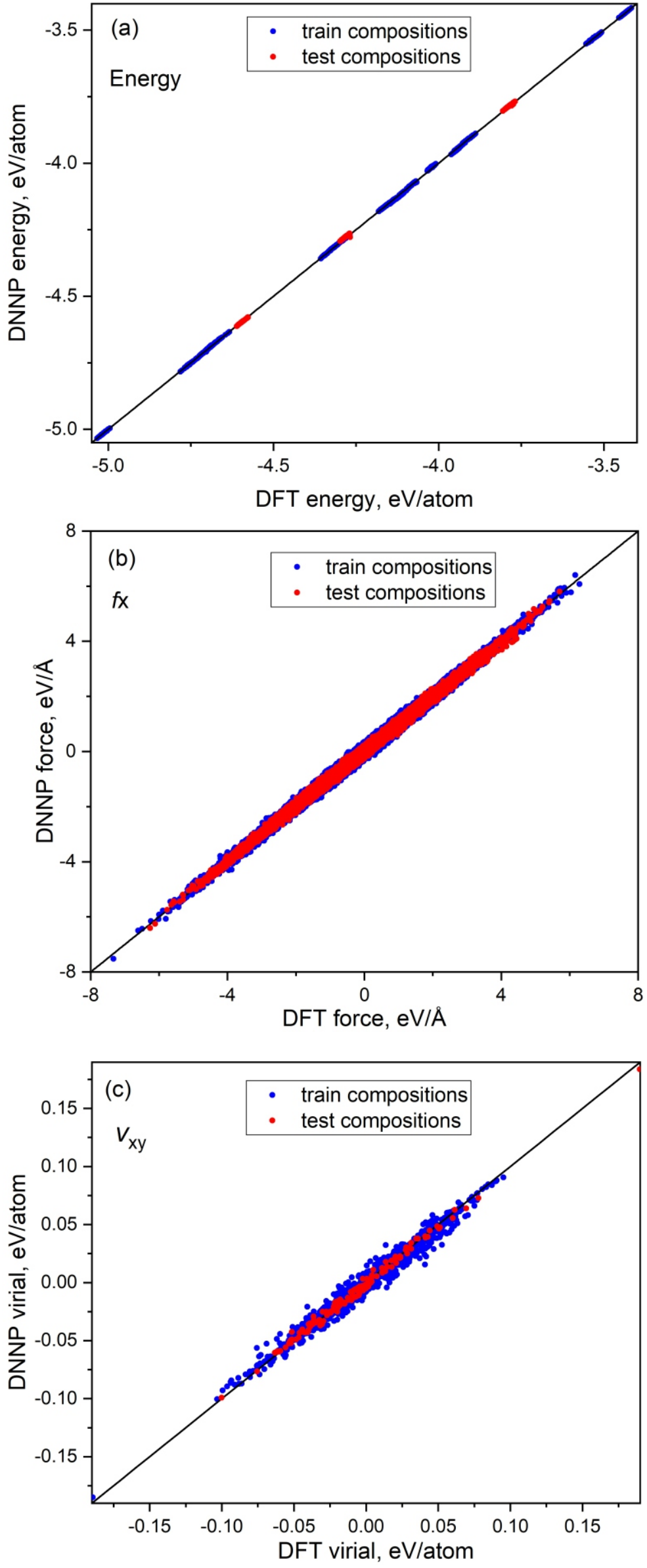}
  \caption{DNNP vs DFT curves for energies, forces and virials in Al-Cu-Ni melts. Blue/red points represent data obtained on compositions included/not included in the training dataset (Fig.~\ref{fig:ternary_plot}).}
  \label{fig:correlation}
\end{figure}

\begin{figure}
  \centering
  \includegraphics[width=0.8\columnwidth]{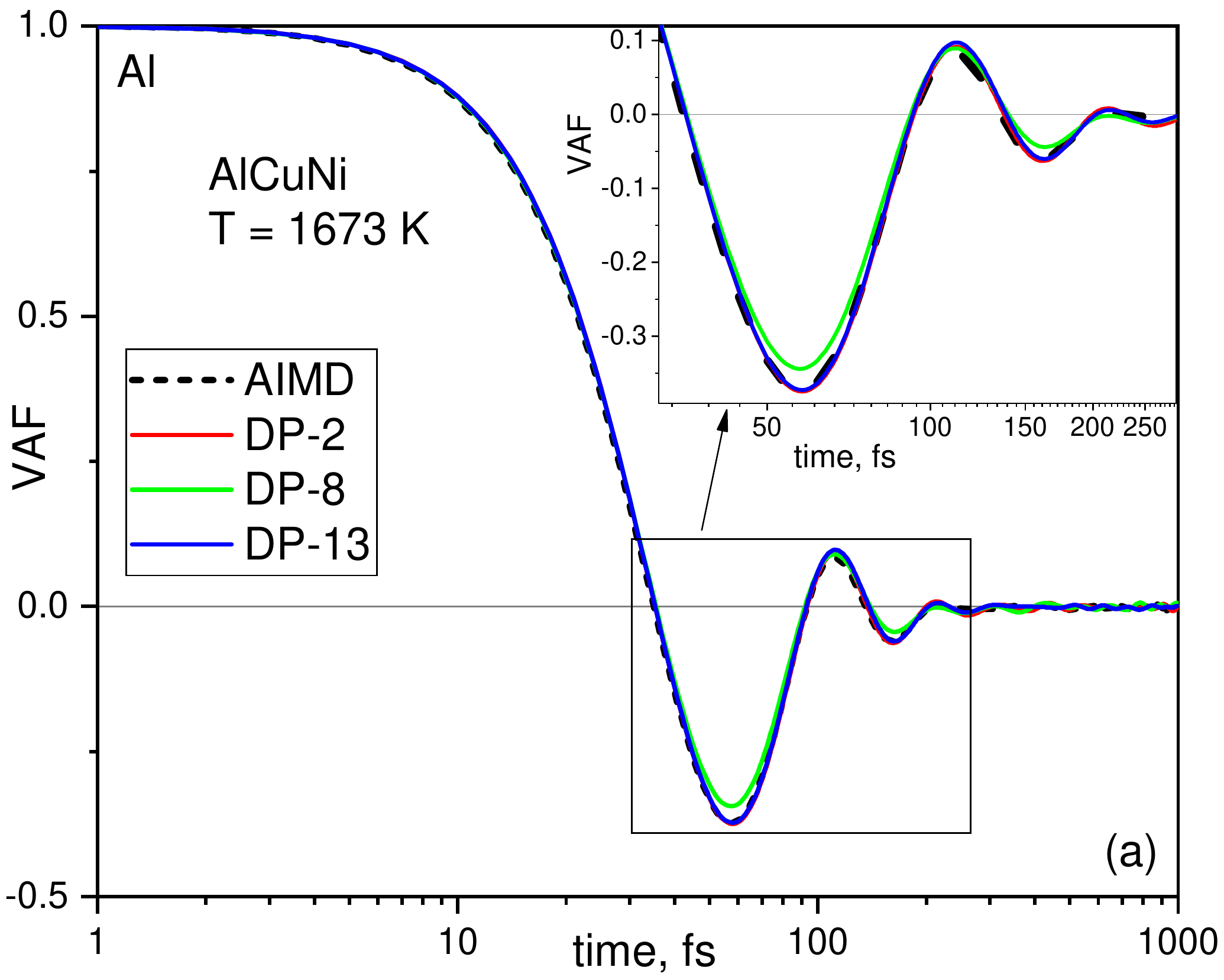}\\
  \includegraphics[width=0.8\columnwidth]{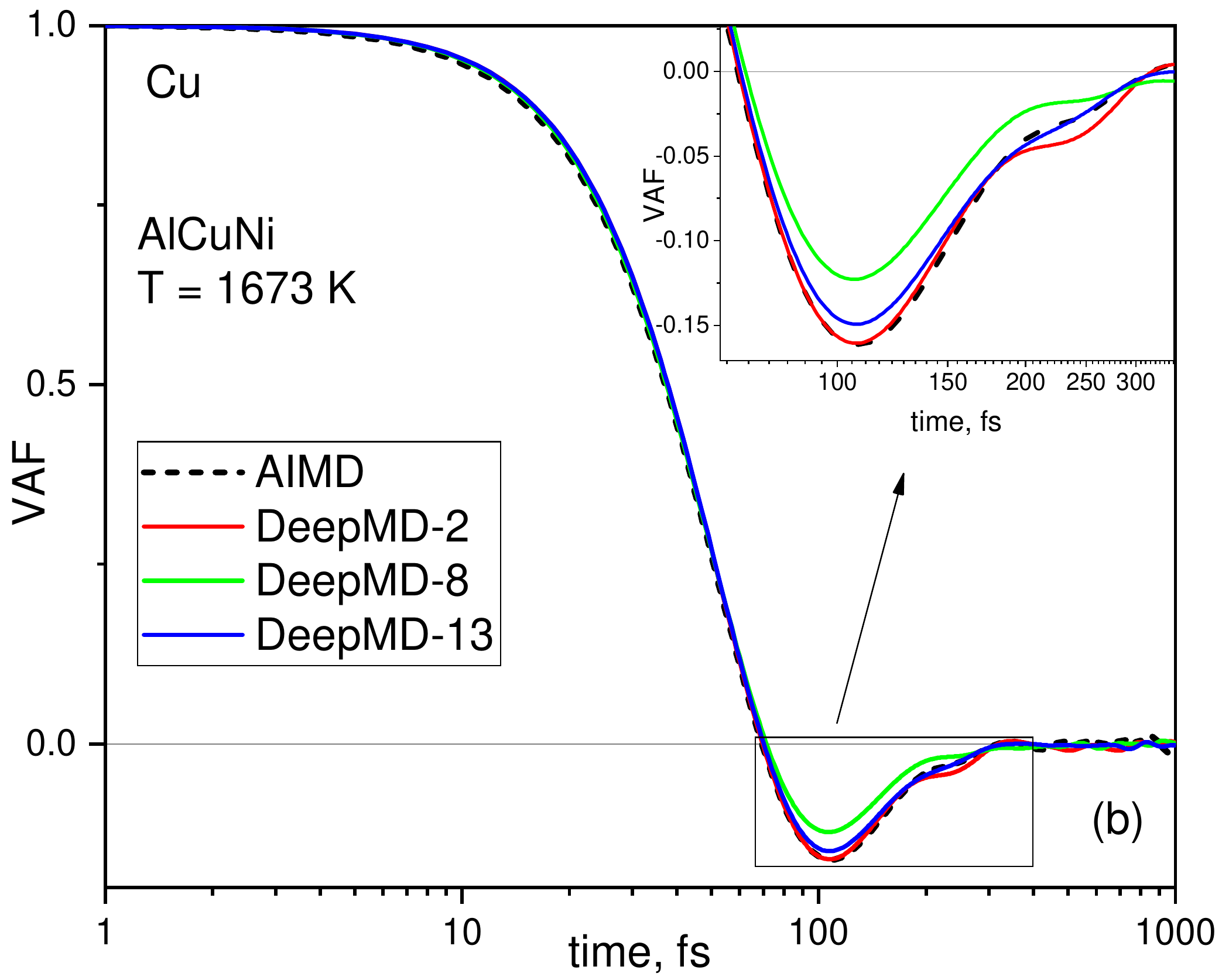}\\
  \includegraphics[width=0.8\columnwidth]{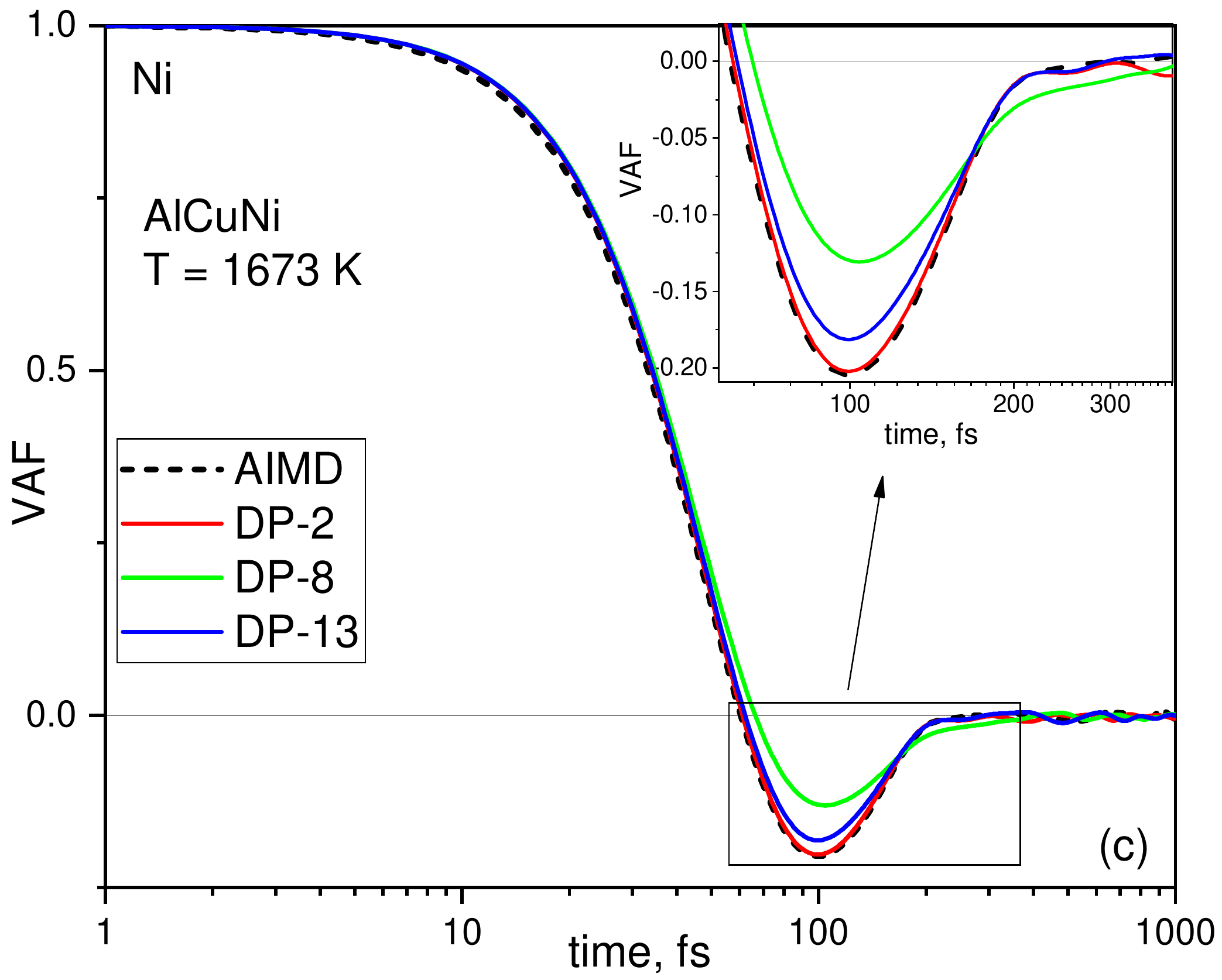}\\
  \caption{ Velocity autocorrelation functions for equatomic AlCuNi melt at $T=1673$ K extracted from AIMD simulations (black dashed lines) as well as from DNNP simulations with three different models (Tab.~\ref{tab:model_parameters}) (colored solid lines).}
  \label{fig:vaf_test}
\end{figure}

\subsection{The search for an optimal DNNP \label{sec:param}}

\begin{figure*}
  \centering
  \includegraphics[width=0.23\textwidth]{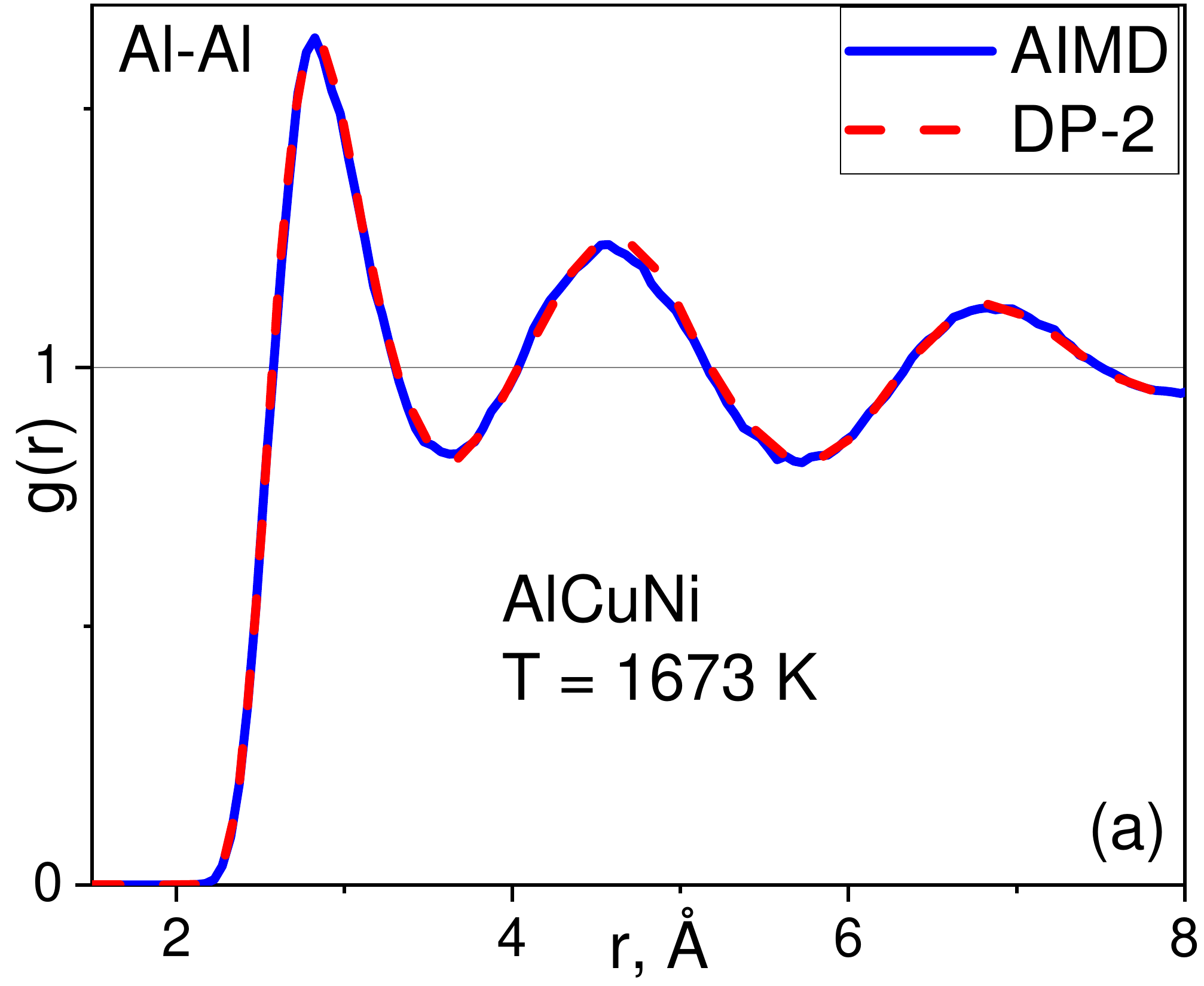}  \includegraphics[width=0.23\textwidth]{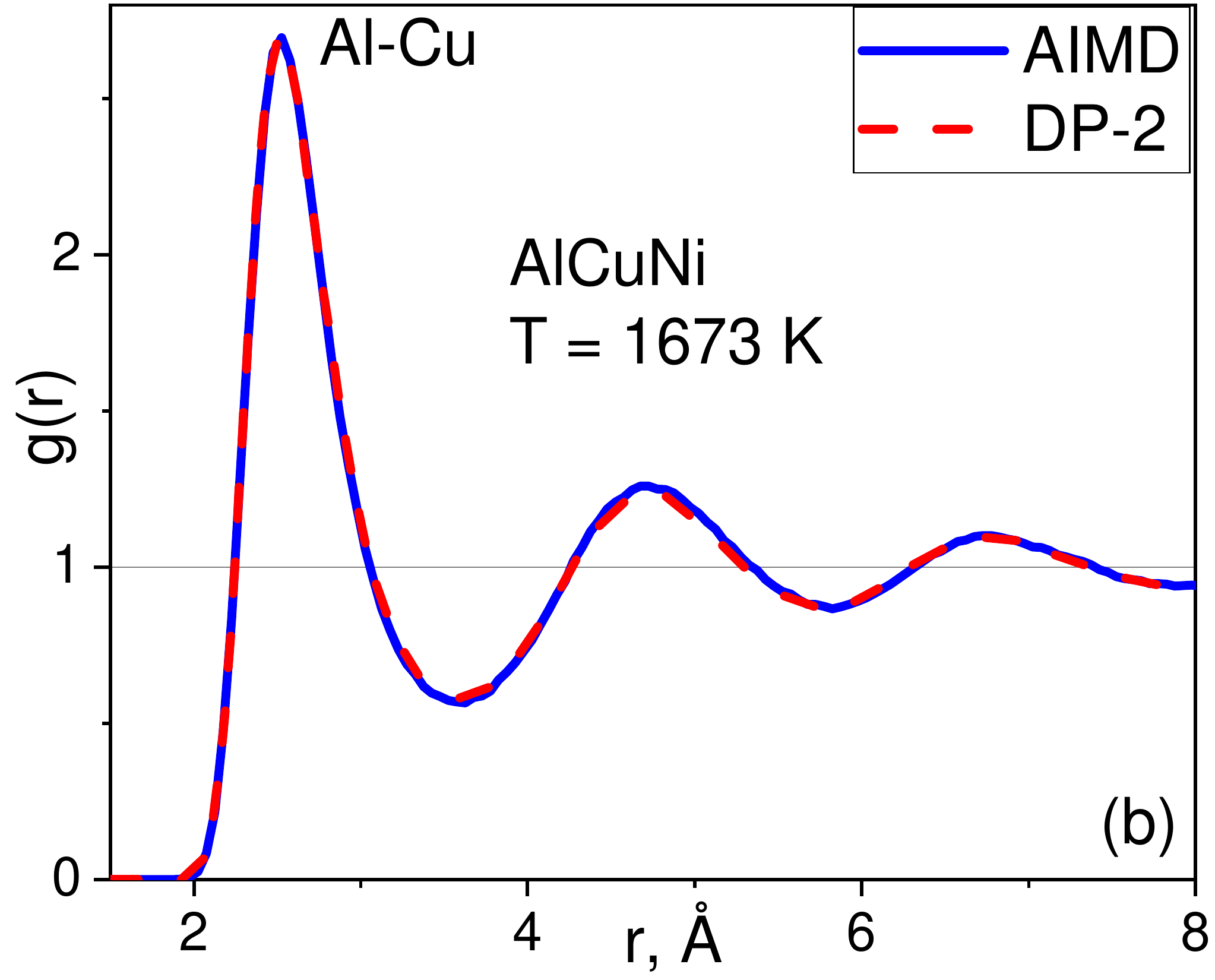}  \includegraphics[width=0.23\textwidth]{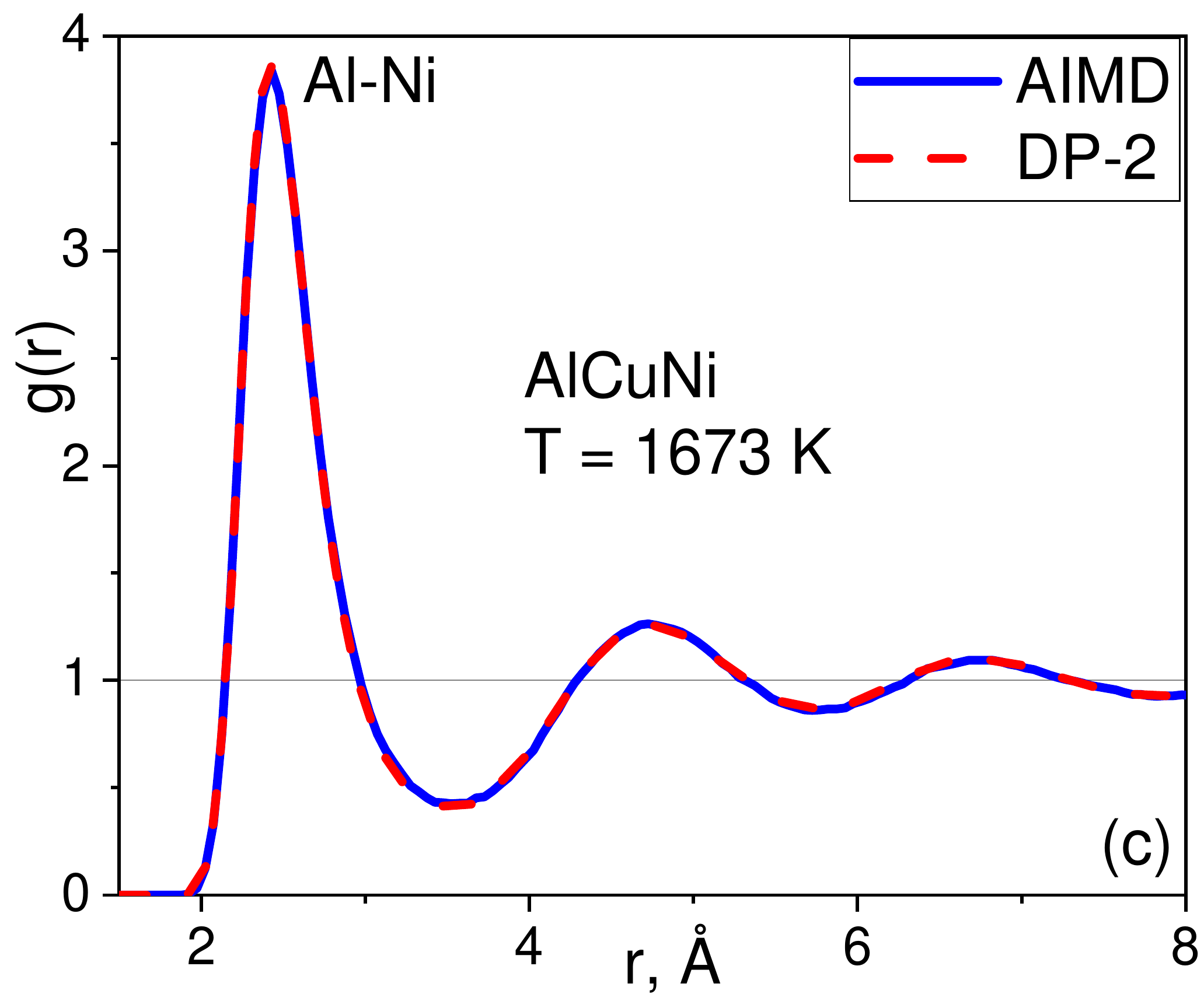}  \includegraphics[width=0.23\textwidth]{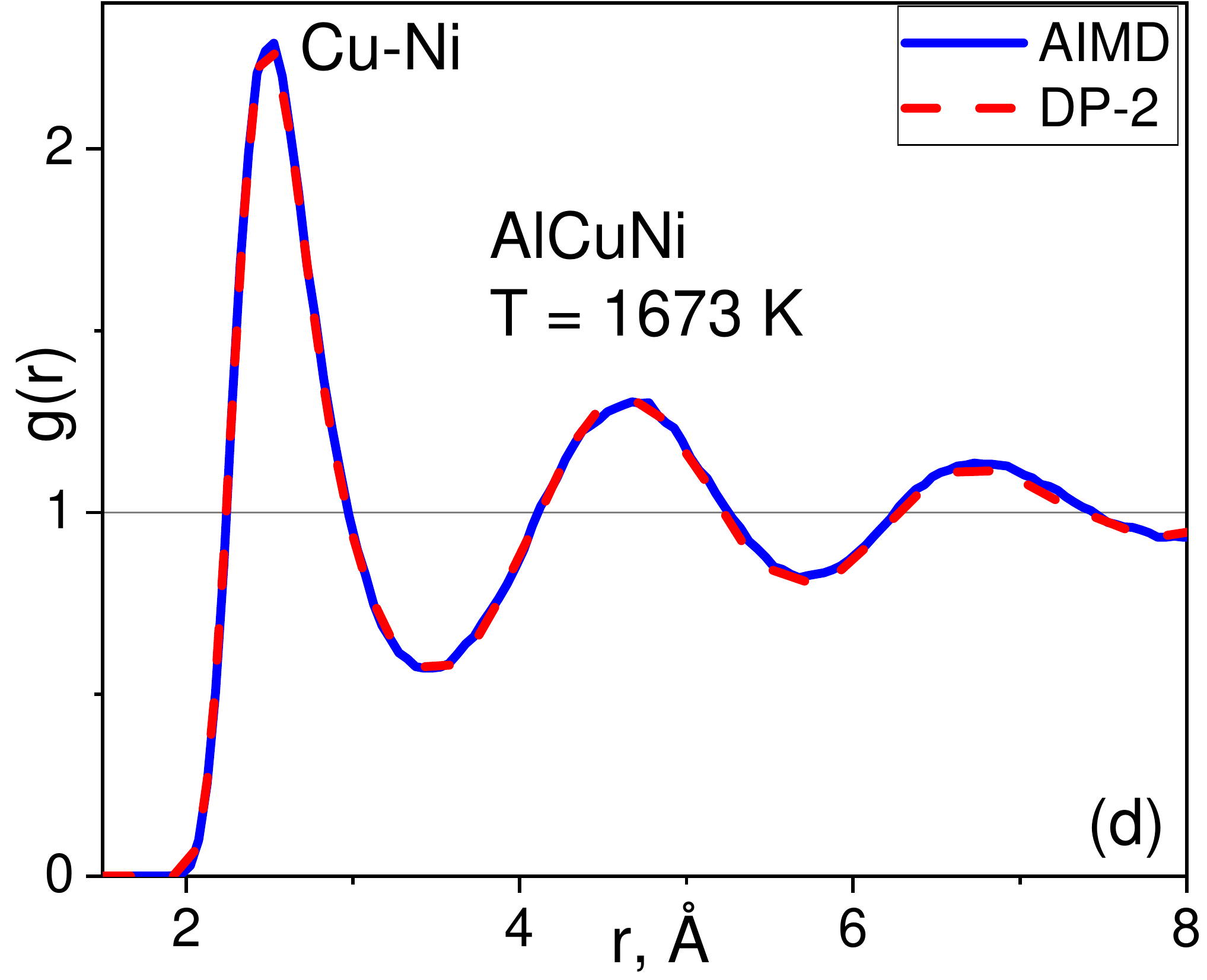}\\
  \includegraphics[width=0.23\textwidth]{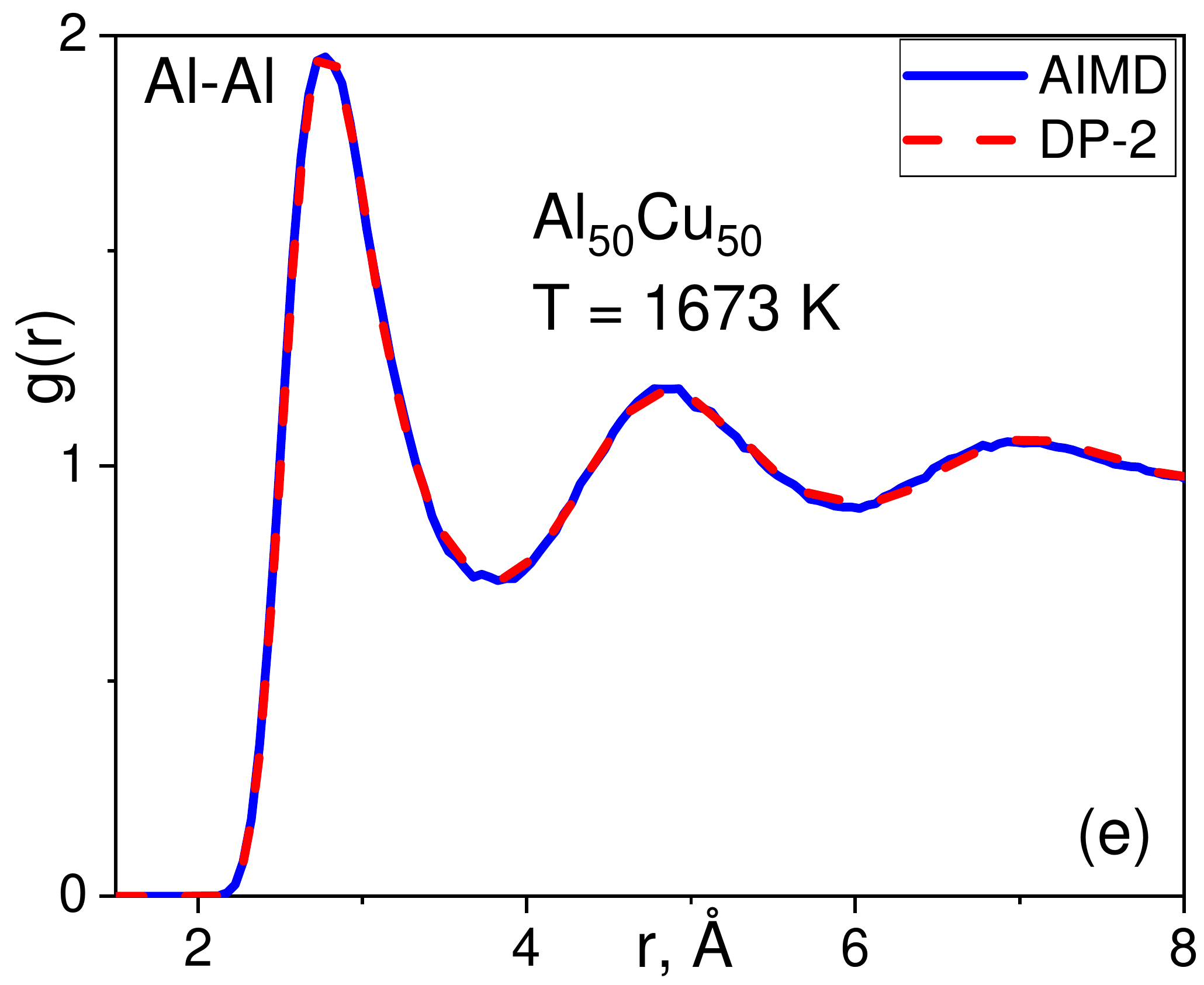}  \includegraphics[width=0.23\textwidth]{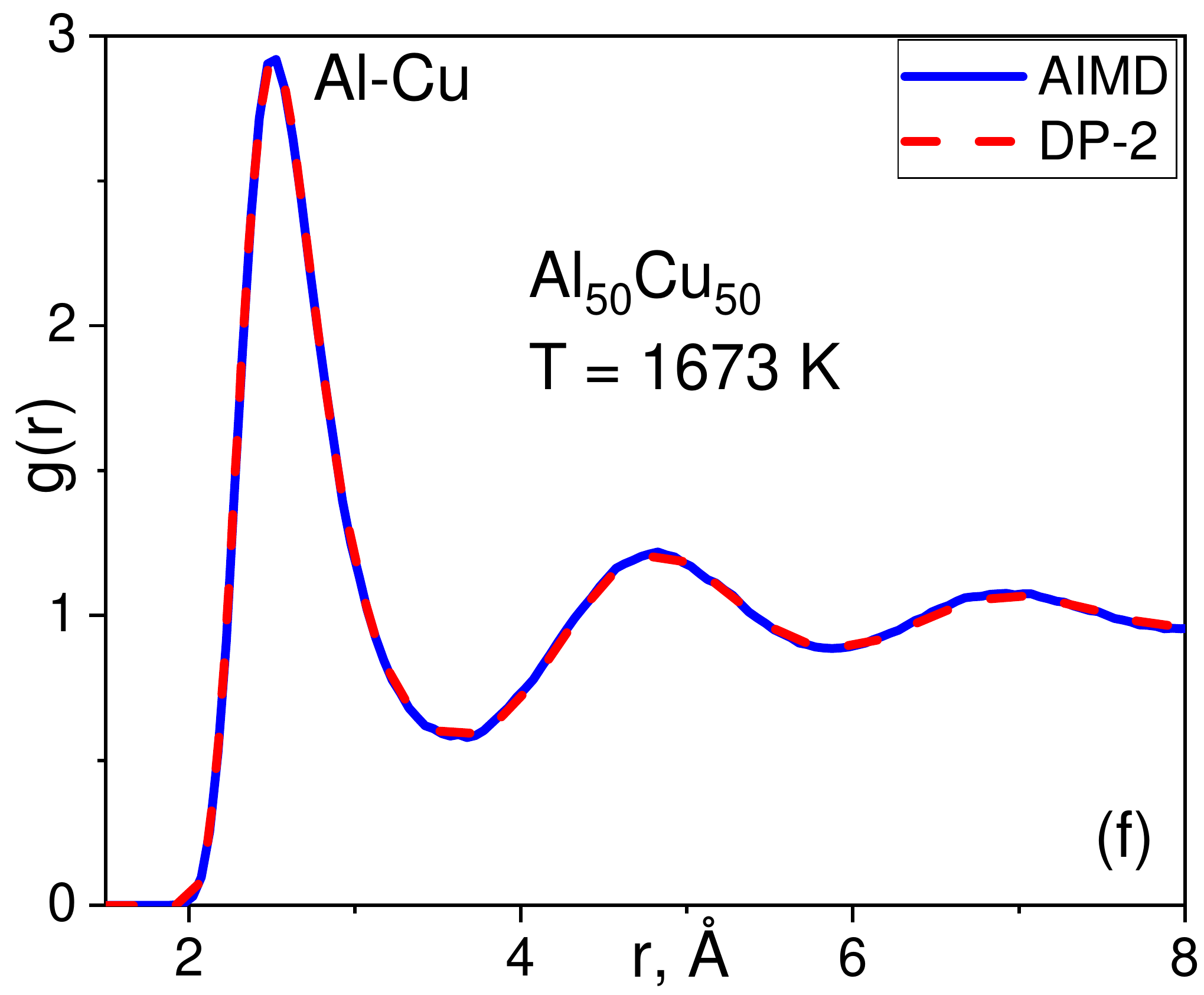}  \includegraphics[width=0.23\textwidth]{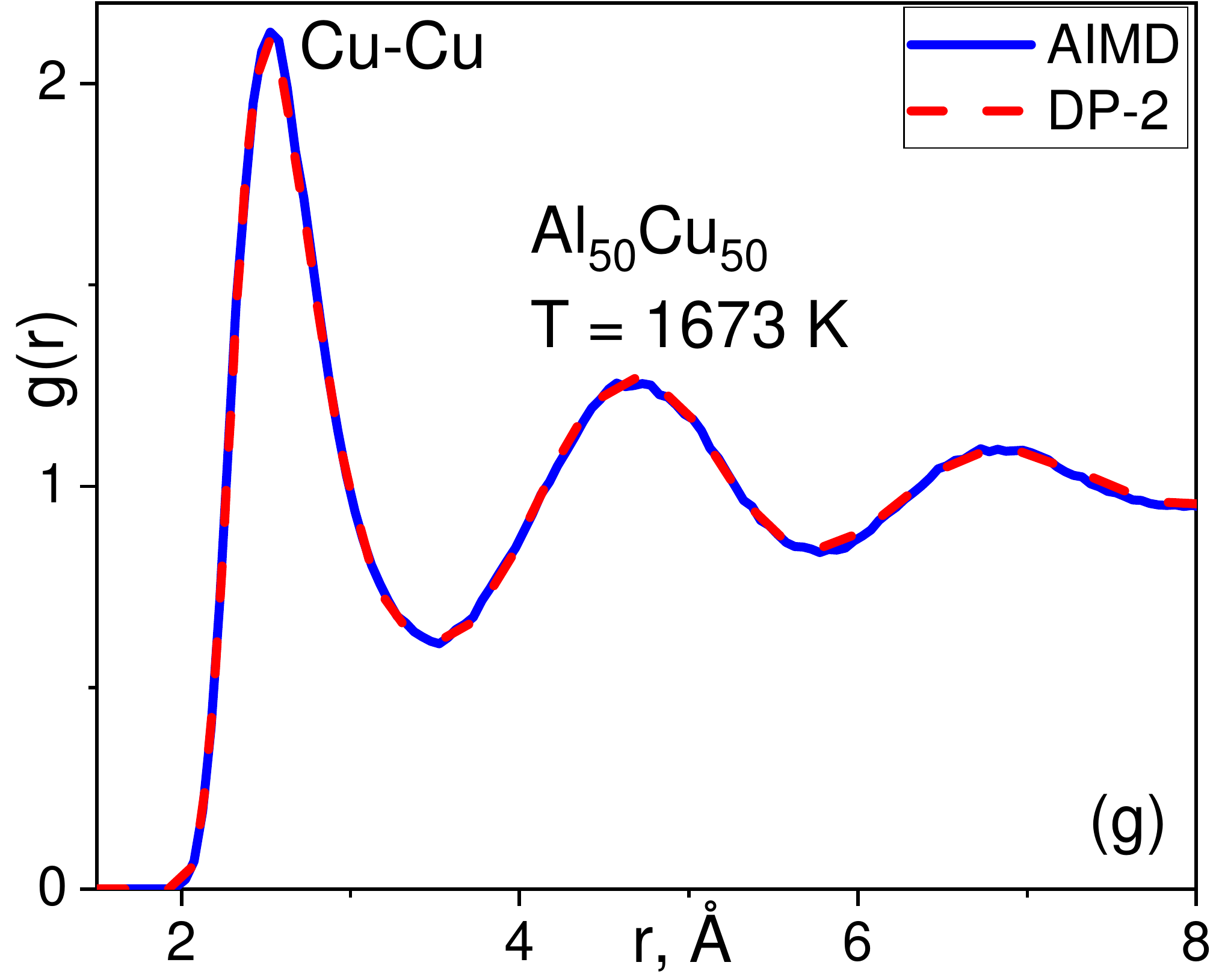} \includegraphics[width=0.23\textwidth]{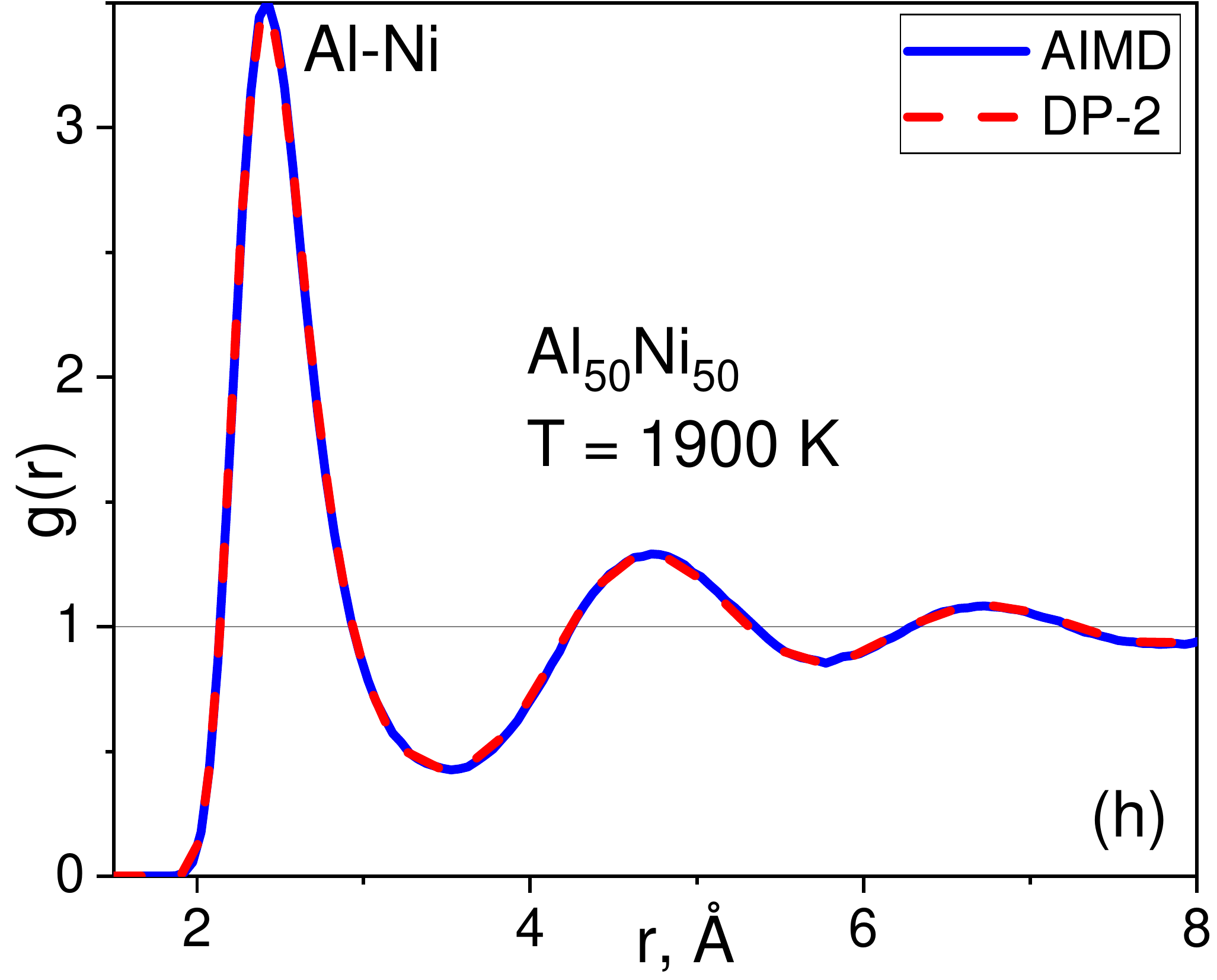} \\
  \includegraphics[width=0.23\textwidth]{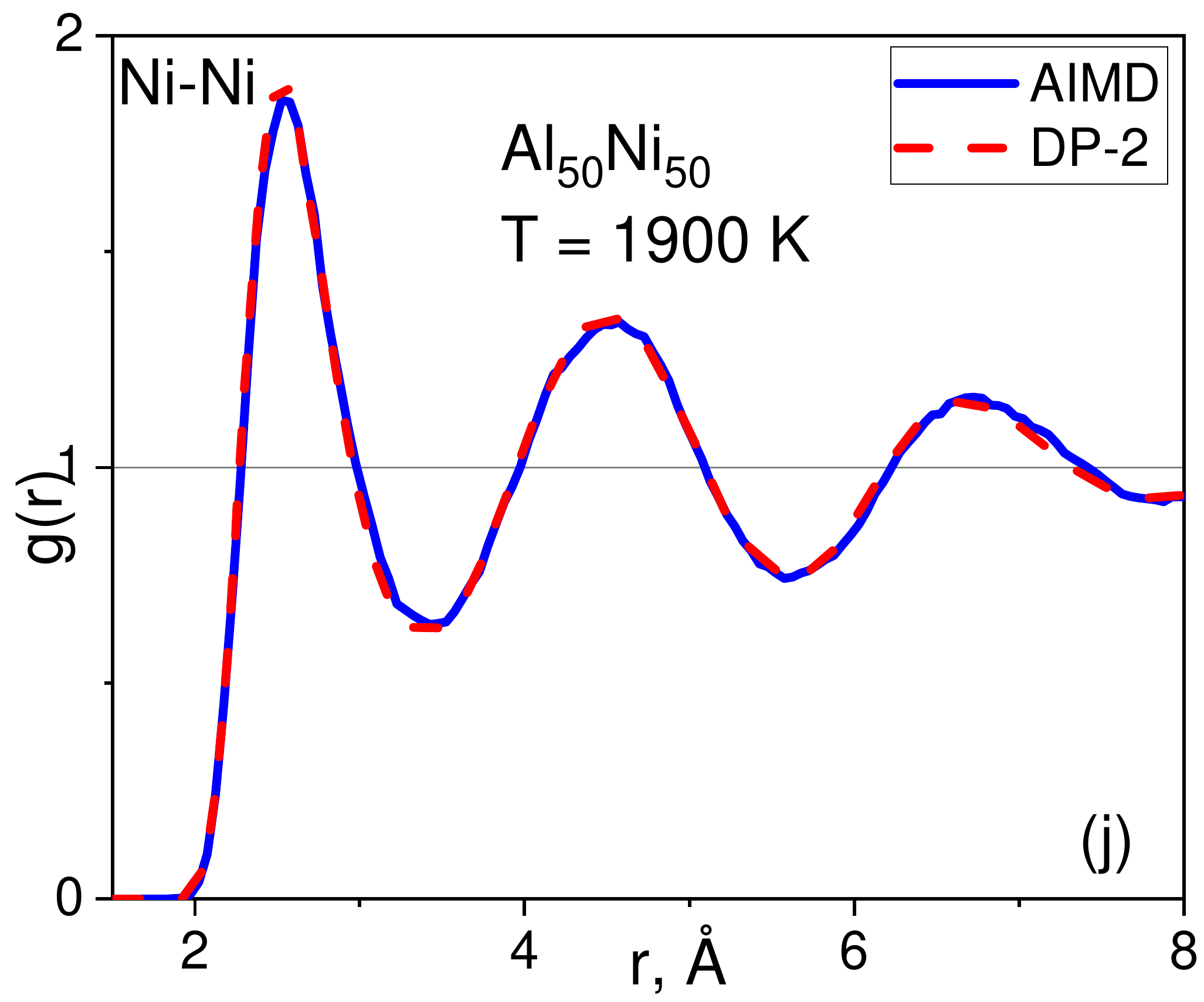}  \includegraphics[width=0.23\textwidth]{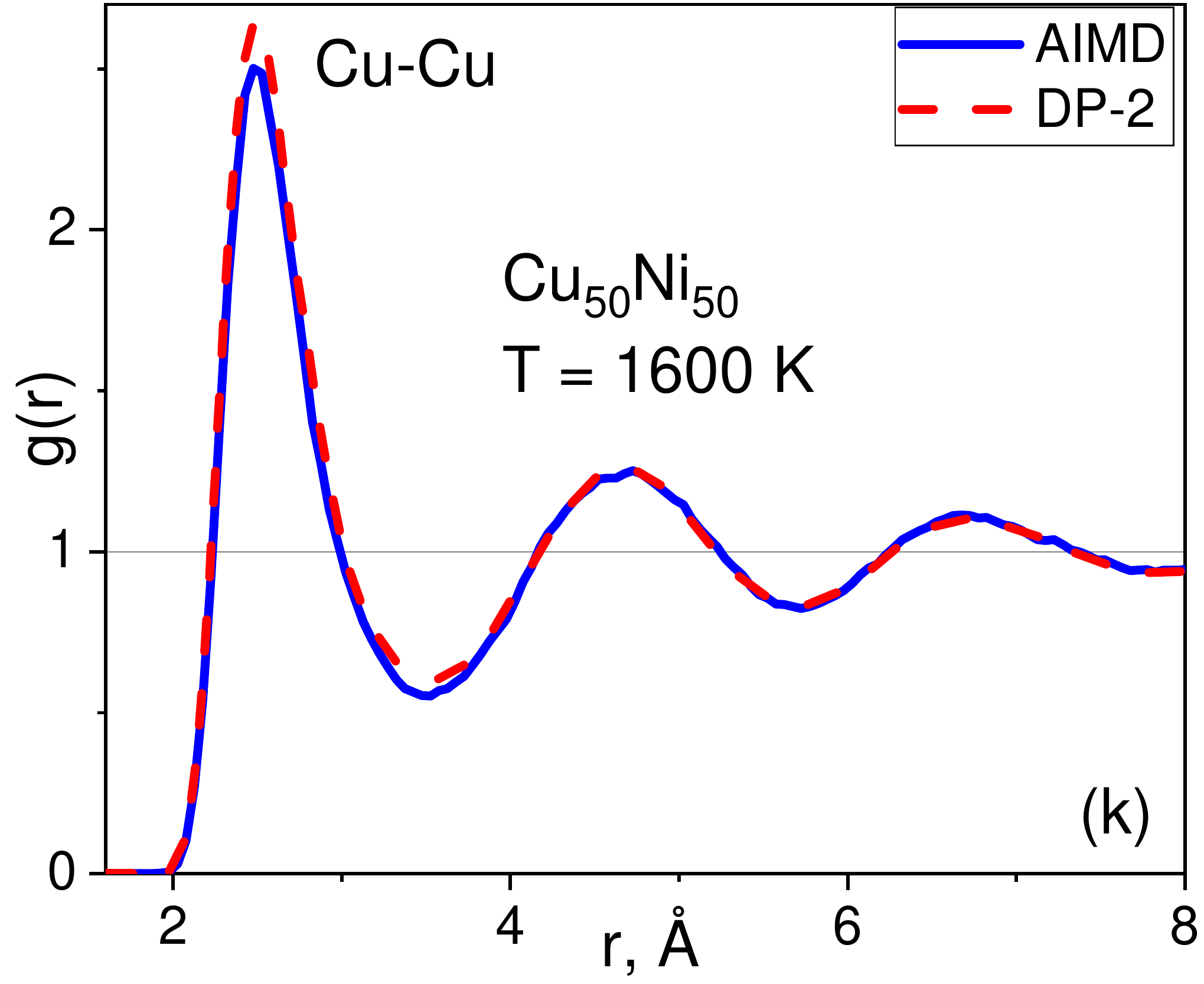} \includegraphics[width=0.23\textwidth]{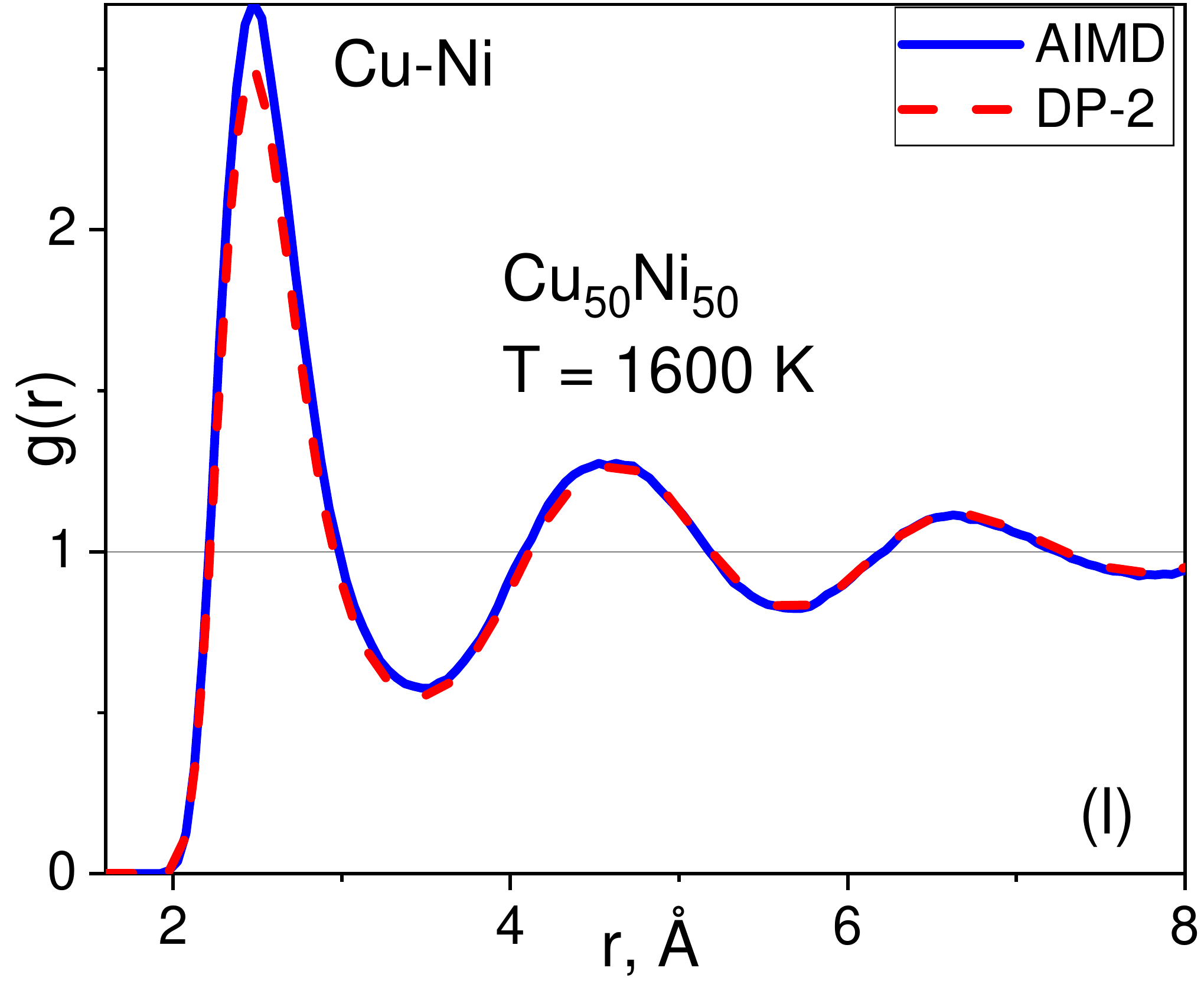} \includegraphics[width=0.23\textwidth]{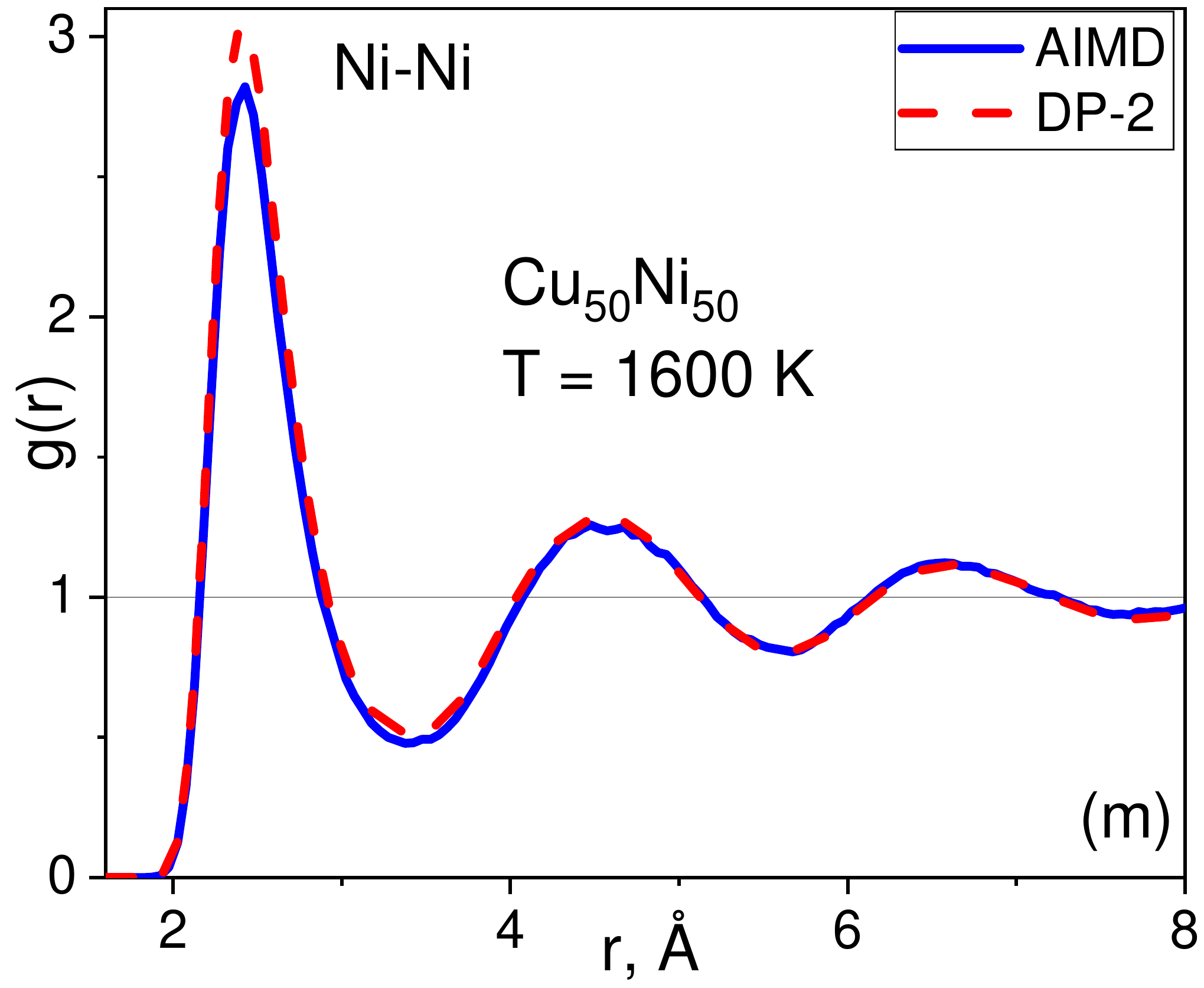}
  \caption{Partial radial distribution functions for Al-Cu-Ni melts at different compositions extracted from AIMD simulations (solid blue lines) as well as from DNNP simulations (red dashed lines).}
  \label{fig:rdfs_aimd}
\end{figure*}

We start with the discussion of DNNPs that are better suited for simulating multicomponent metallic melts. As we mentioned above, the number of possible variants of neural networks hyperparameters and learning scheme parameters is enormous and so we do not hope to find the best model. Moreover, such an optimal DNNP configuration is expected to depend on the training dataset. Thus, our purposes are to determine the parameters that have a major impact on both the accuracy and computational efficiency of DNNPs and then roughly localize an optimal domain of their values for Al-Cu-Ni melts. To do that, we chose a base model (see model-1 in Tab.~\ref{tab:model_parameters}) whose small variations are successfully used in the literature for describing unary and binary liquids, glasses, and crystals~\cite{Wen2019PRB,Andolina2020JCP,Gartner2020PNAS}. When we design a series of DNNPs by simplifying/complicating the embedding and/or fitting networks of the base model as well as considering different prefactors of the loss function and model cutoff radii. All the designed models were trained on the dataset described above (see Fig.~\ref{fig:ternary_plot}) for 400,000 iterations until the components of the loss function corresponding to energy, forces and virials demonstrate the saturation of the training process. We use a small batch size of 2. Taking into account that the number of configurations is about 30,000, the entire training dataset is passed about 25 times during the training procedure. The resulted DNNPs will be further referred to as DP-$n$, where $n$ is the number of the model in Tab.~\ref{tab:model_parameters}.

We perform a preliminary analysis of the accuracy and computational efficiency of the models developed. To estimate the accuracy, we calculate the root mean square errors (RMSEs) for energies, forces, and virials calculated by DNNP in comparison with those from DFT calculations. The units of RMSEs are meV/atom for energies and virials and meV/{\AA} for forces. The computational performance is expressed in ns/day for a test simulation of 4096-particle AlCuNi alloy performed on a single GPU. The results of models evaluation are presented in Tab.~\ref{tab:model_parameters}. By analyzing the results, we draw several important conclusions, which are discussed below.

\begin{figure}
  \centering
  \includegraphics[width=0.8\columnwidth]{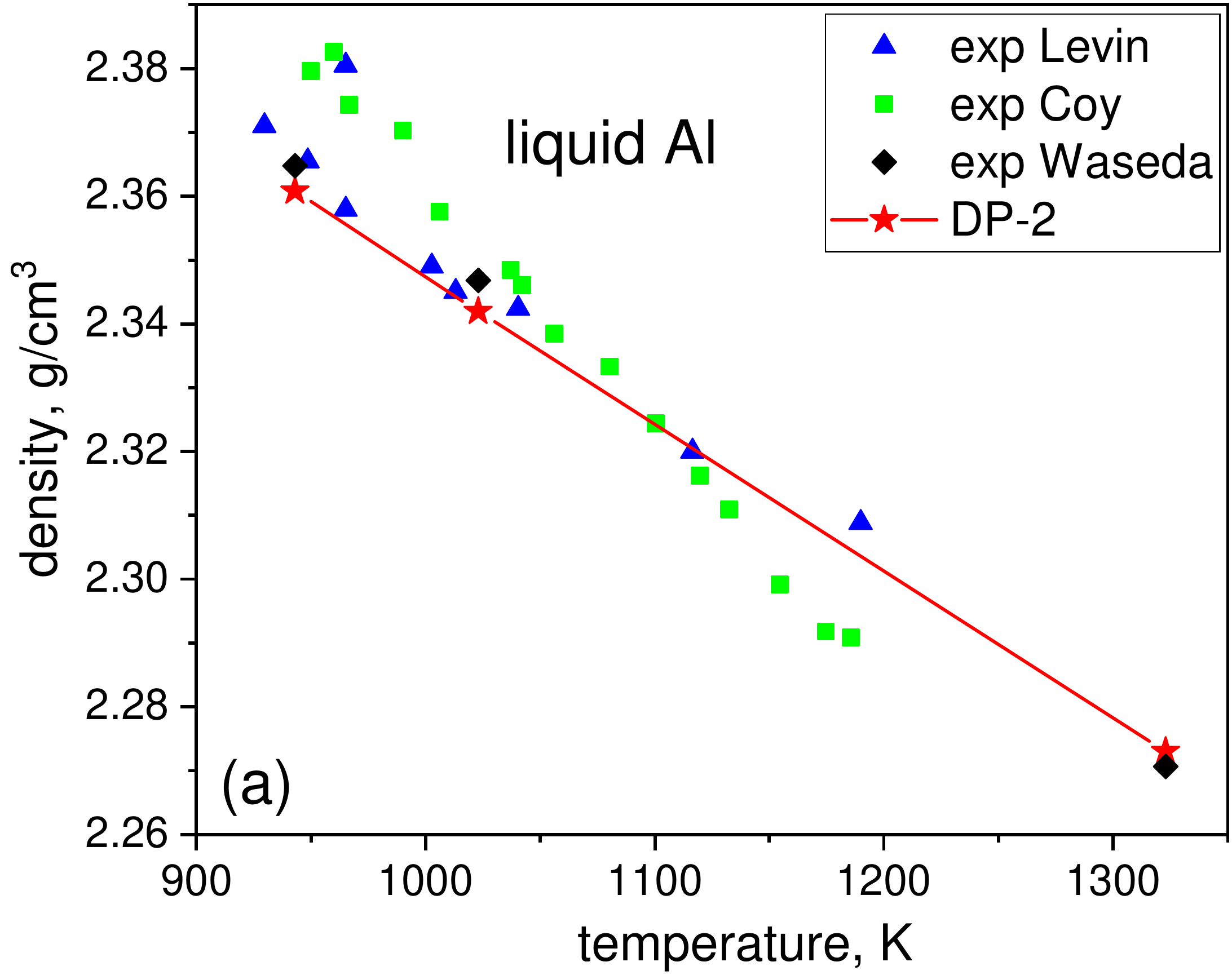}\\
  \includegraphics[width=0.8\columnwidth]{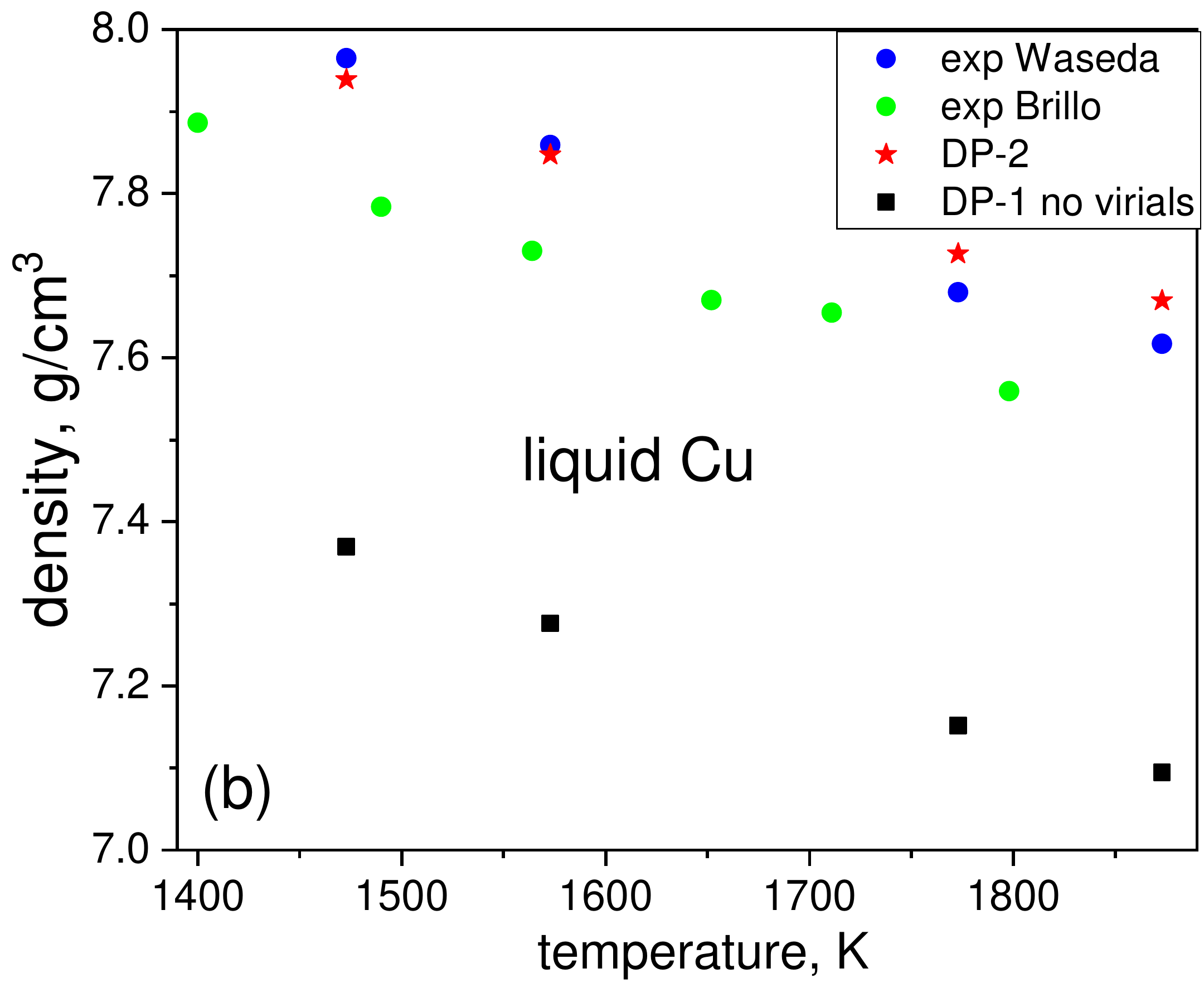}\\
  \includegraphics[width=0.8\columnwidth]{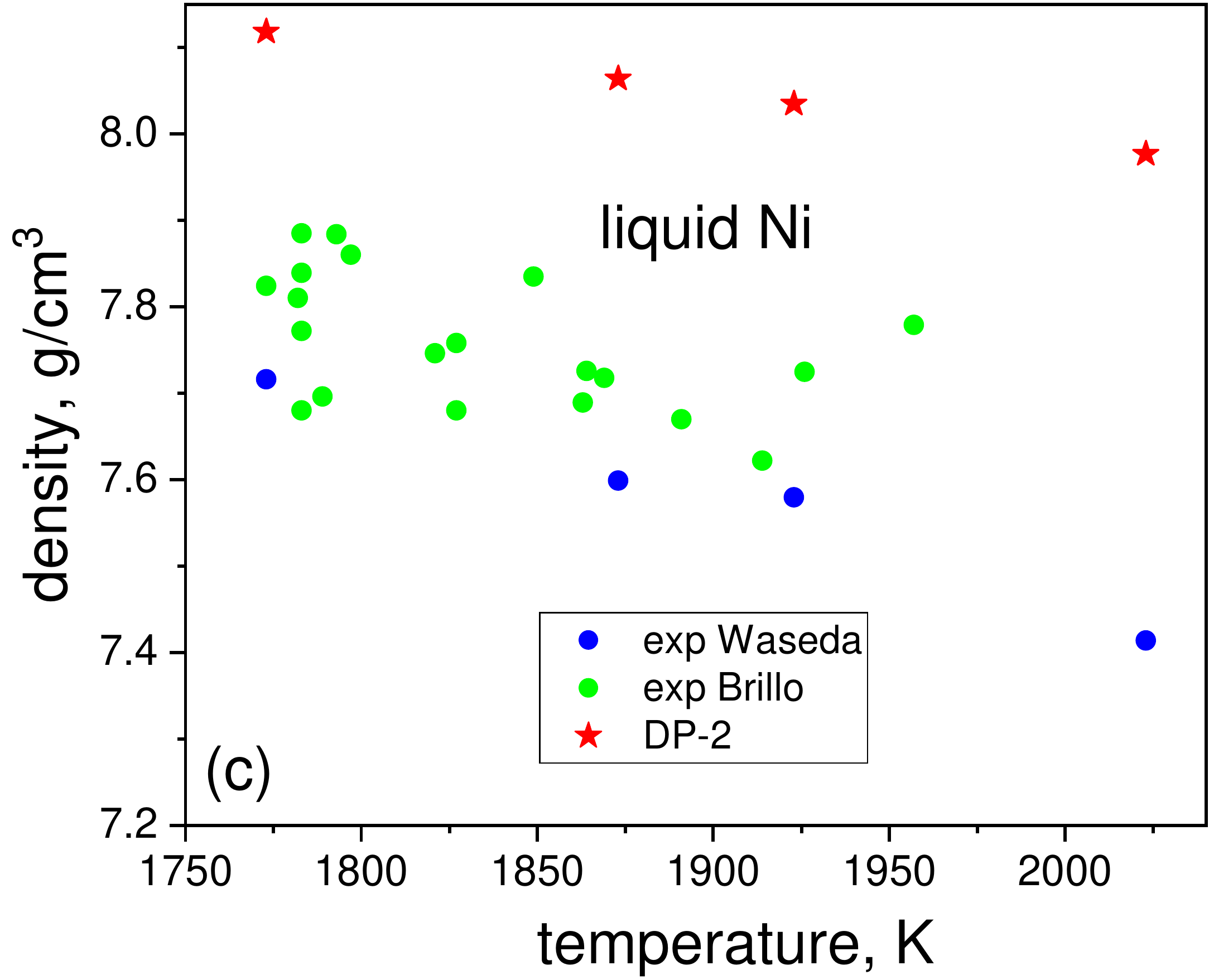}
  \caption{ Temperature dependencies of equilibrium density of liquid Al, Cu, and Ni. Symbols represent experimental data taken from~\cite{Waseda1980textbook,Brillo2003IntJThermPhys,Assael2006JPhysChemRefDat}. Red stars are the results of DDNP-based simulations.}
  \label{fig:dens}
\end{figure}

We start with the discussion of the learning scheme, which is the choice of start and limit values of the pferactors in the loss function corresponding to energies, forces, and virials. We have tested different variants of the learning schemes and concluded that all of them lead to more or less the same final results.   However, we have made an important conclusion regarding the importance of virials in the training procedure. Virials are often excluded from the learning scheme putting $p_v^{\rm start}=p_v^{\rm limit}=0$. We guess that this is due to an assumption that virials are determined by the forces and so including the latter is enough. To check this point we trained two models whose only difference was the values of virials prefactors: $p_v^{\rm start}=p_v^{\rm limit}=0$ for the DP-1 and $p_v^{\rm start}=p_v^{\rm limit}=0.1$ for the DP-2 (see Tab.~\ref{tab:model_parameters}). We see that both models demonstrate similar accuracy in describing energies and forces but crucial difference in describing virials. We will see below that this difference is crucial for obtaining an equilibrium density in NPT simulations (see Fig.~\ref{fig:dens}b). Thus, we will further use the learning scheme $(0.2, 0.2, 500, 500,0.1,0.1)$ where virials are included and all the prefactors are constant during the training.

We find that both the accuracy and the efficiency of DNNP depend strongly on the structure of the embedding net; the impact of the fitting net structure is much less. Indeed, radical variation of the fitting net complexity from (50,50) to (240,240,240,240), at the fixed parameters of the embedding net, does not change essentially the quality of DNNP (compare DPs 2,4,5). At the same time, simplification of the embedding net (with respect to the base model) at fixed fitting net, make the model substantially less accurate but more efficient (compare DPs 6,7,8).

We also observe that variations of the cutoffs in comparison with the case of $(5,6)$ do not essentially change the results if the structure of the embedding net is fixed. At the same time, an increase of the $r_{\rm c}$ up to 7 {\AA} accompanied with an increase of embedding net complexity cause a slight increase of the accuracy but also a noticeable decrease in the computational efficiency (see DP-13). Interestingly, that going further in this way we do not gain in accuracy but lose drastically in performance (see DP-14). The probable reason is that complex models are more prone to overfitting and so larger datasets are needed to train them properly.

 For the DP-2, we present in Fig.~\ref{fig:correlation} the DNNP vs DFT plots for energy, forces, and virials, which demonstrate how closely the values predicted by DNNP correlate with DFT data. We see that DP-2 provides distinct linear correlations for all the quantities. Other models reveal similar behavior.

\begin{figure}
  \centering
  \includegraphics[width=0.76\columnwidth]{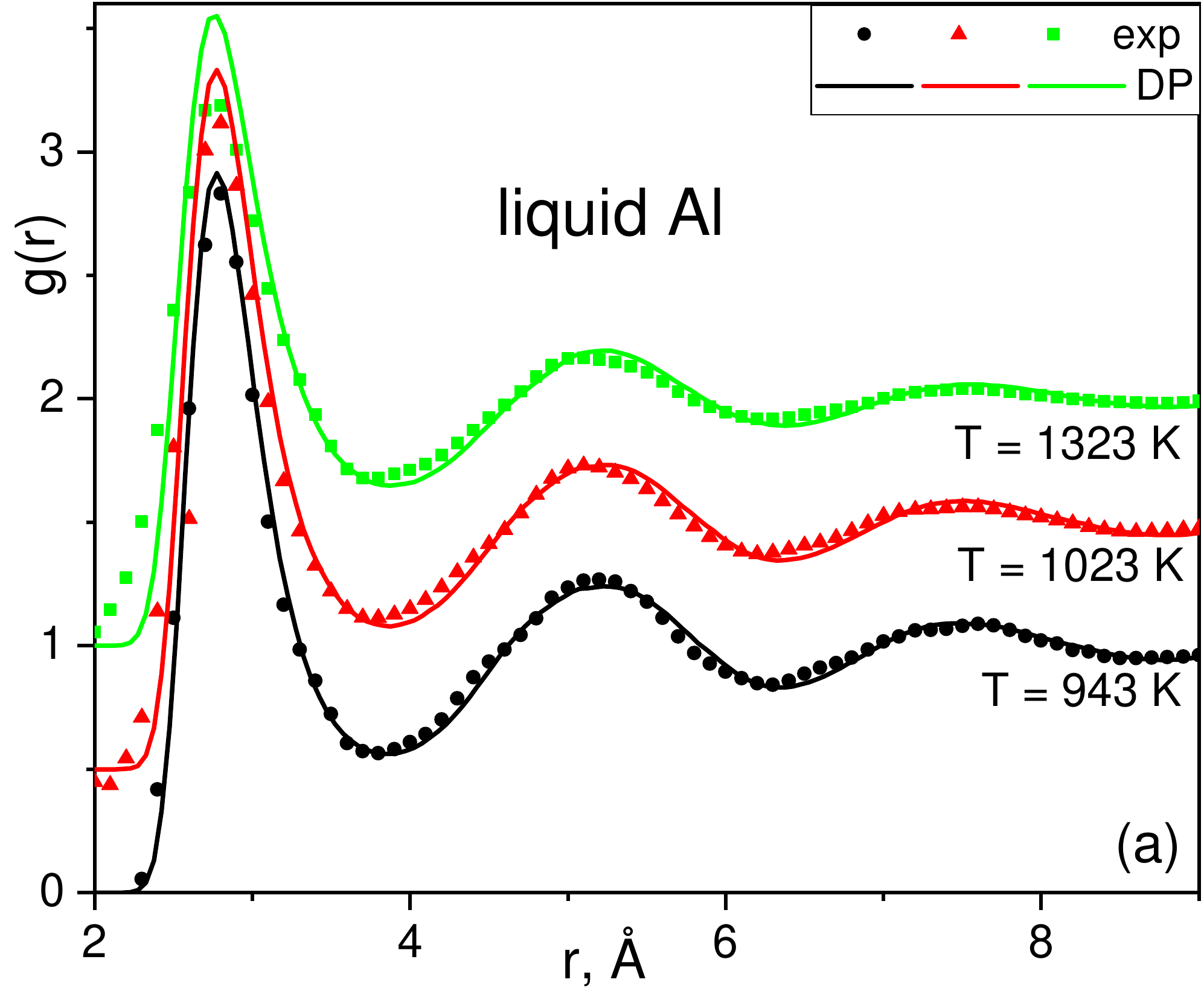}\\
  \includegraphics[width=0.76\columnwidth]{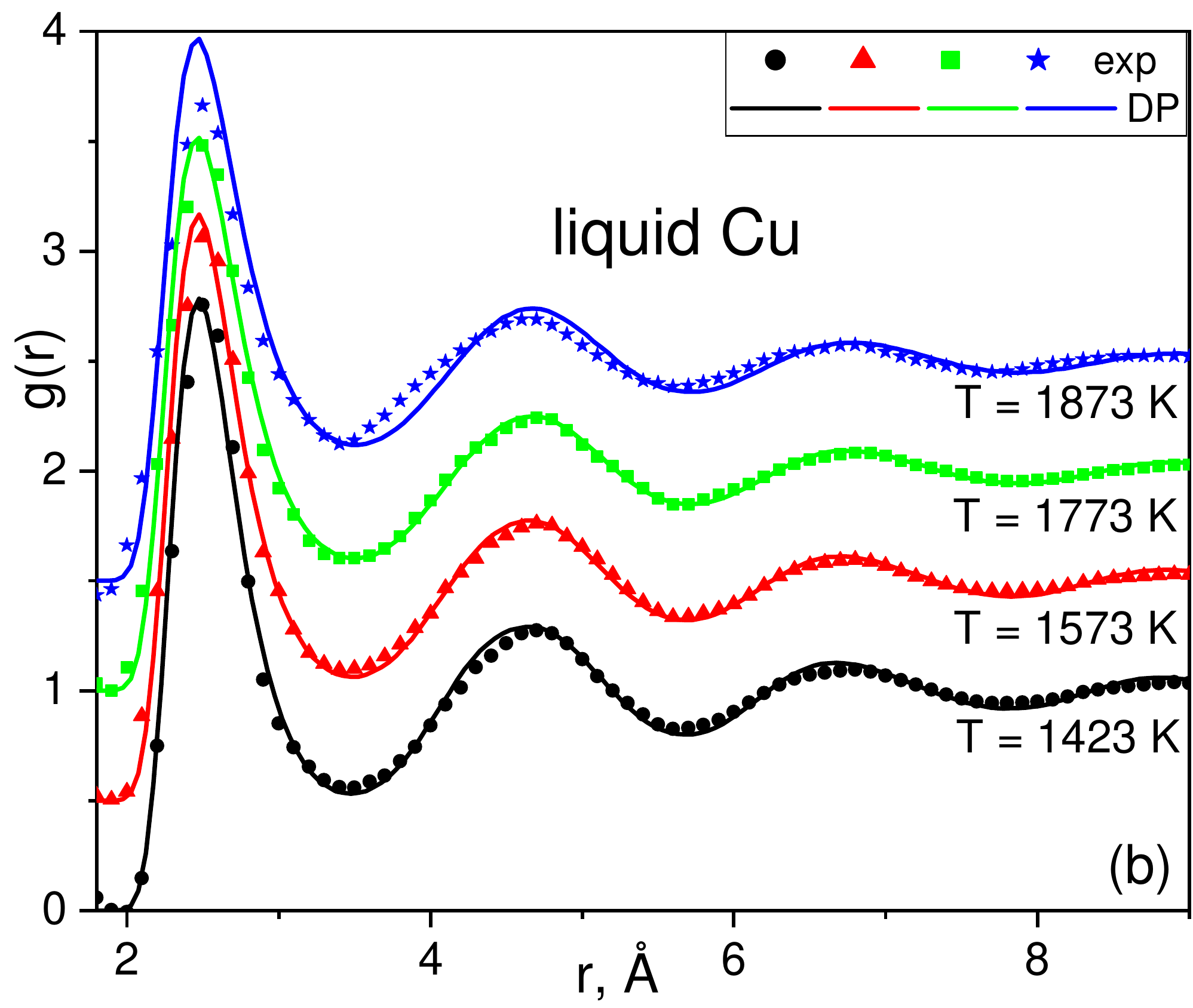}\\
  \includegraphics[width=0.76\columnwidth]{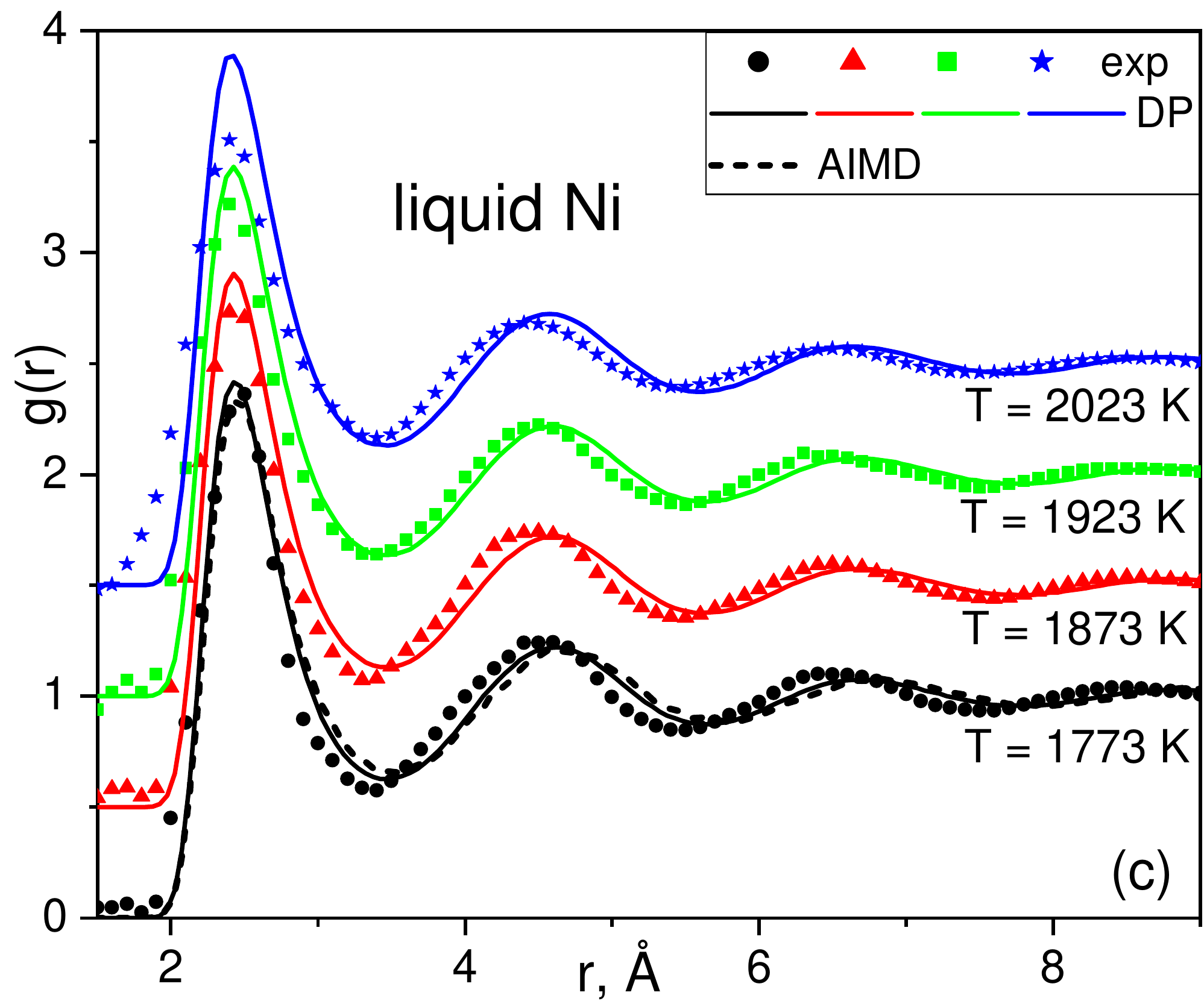}
  \caption{ Radial distribution functions for liquid Al, Cu, and Ni at different temperatures. Symbols represent experimental data taken from~\cite{Waseda1980textbook}. Solid lines are the results of DDNP-based simulations. The curves at elevated temperatures are shifted along Y-axis for visual clarity.}
  \label{fig:RDF_exp}
\end{figure}

Thus, analyzing Tab.~\ref{tab:model_parameters},  we choose three DNNPs to further testing and comparing: DP-2 as a well-balanced model, DP-13 as the most accurate model in respect to RMSE, and DP-8 and the most efficient one. To compare these models more carefully, we calculate velocity autocorrelation functions (VAFs) for equiatomic AlCuNi melt using corresponding DNNPs and compare the results with those obtained by AIMD (Fig.~\ref{fig:vaf_test}). The analysis of VAFs is a good test for interatomic potentials because dynamical correlation functions are very sensitive to the accuracy of the description of interatomic forces and energies. We see from the figure that DP-2 provides the best agreement with AIMD data. The DP-8 expectably demonstrates the worst result among the three models, although the general agreement is satisfactory. Surprisingly, DP-l3, which has the lowest RMSE, reveals slightly worse VAF accuracy than DP-2, especially for Cu and Ni.

 Thus, DP-2 provides an optimal accuracy/performance ratio and so we consider it as the main model. Below, we will test this model carefully by calculating observable properties at different compositions in comparison with both AIMD simulations and experimental data.

 It should be noticed that even DP-8, which has the highest RMSEs, reveals reasonable accuracy in describing structural and dynamical properties of AlCuNi melts (see Fig.~\ref{fig:vaf_test}). Taking into account that this simple model is four times faster, we conclude that a simplification of DNNPs may be a reasonable strategy in simulations where high computational efficiency is an important requirement.

\subsection{DNNP verification on training compositions}
As we showed above, the developed DNNP (DP-2) reveals small RMSEs between AIMD and DNNP calculations for energy, forces, and virials. So we expect that the potential would provide good accuracy in simulating observable properties of Al-Cu-Ni melts at least at the compositions that were included in the training dataset. The results for VAFs in Fig.~\ref{fig:vaf_test} support this idea. To check this point further, we extract partial radial distribution functions (RDFs) of the melts from both AIMD and DDNP simulations at the training compositions. The total number of different RDF curves is very large so we show in Fig.~\ref{fig:rdfs_aimd} only a few of them. We see excellent agreement between AIMD and DNNP curves. General analysis of the partial RDFs allows drawing some conclusions: (i) character of chemical interaction between certain species can depend on composition (compare, for example, Al-Al RDFs in Figs.\ref{fig:rdfs_aimd} a,e and Ni-Ni RDFs in Figs.\ref{fig:rdfs_aimd} j,m); (ii) Pronounced first RDF peaks suggest strong chemical interaction for Al-Cu and Al-Ni pairs, especially for the latter. These conclusions are in agreement this the results obtained in Ref.~\cite{Kamaeva2020JPCM}.

 \begin{figure*}
  \centering
  \includegraphics[width=0.25\textwidth]{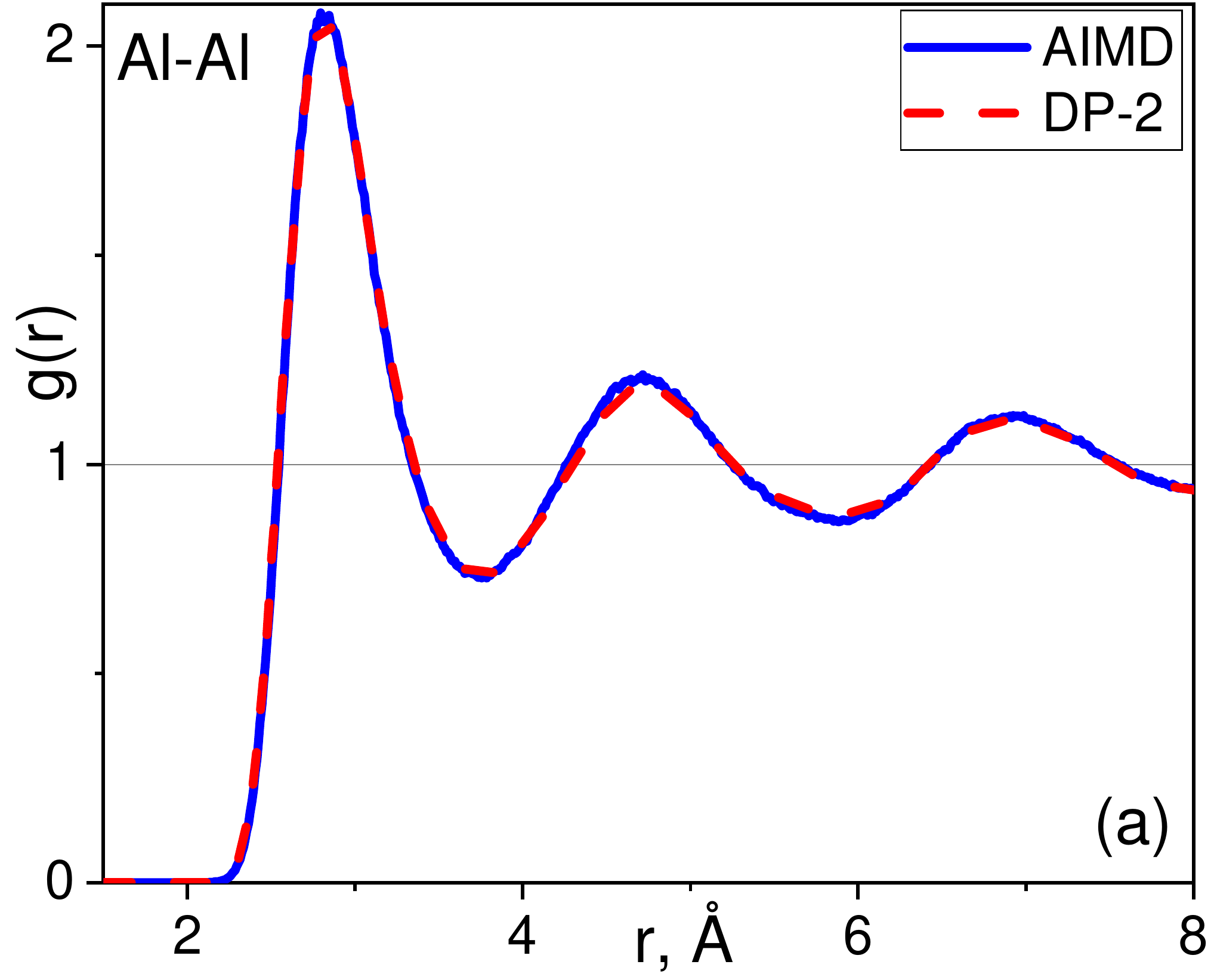}  \includegraphics[width=0.25\textwidth]{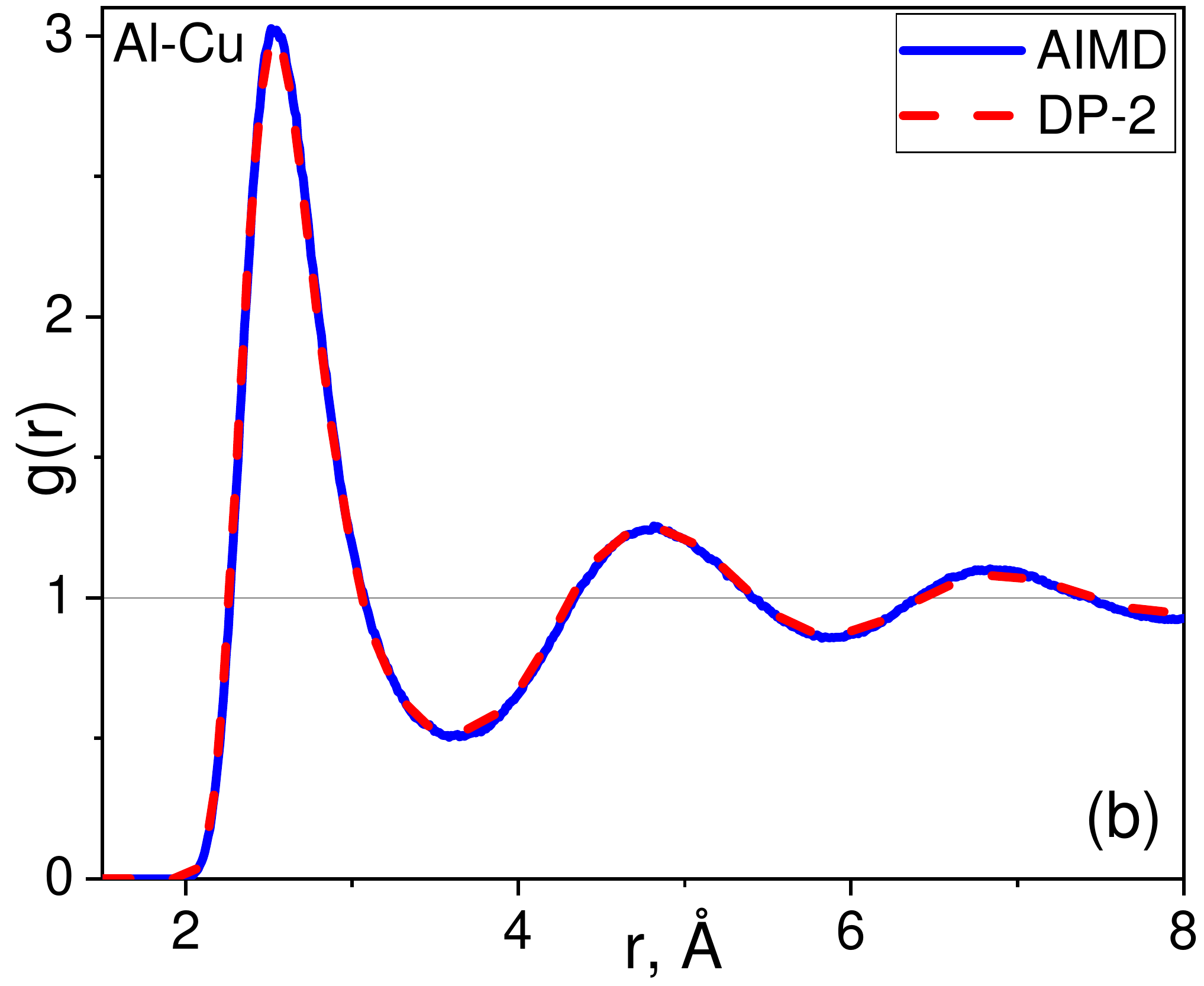}  \includegraphics[width=0.25\textwidth]{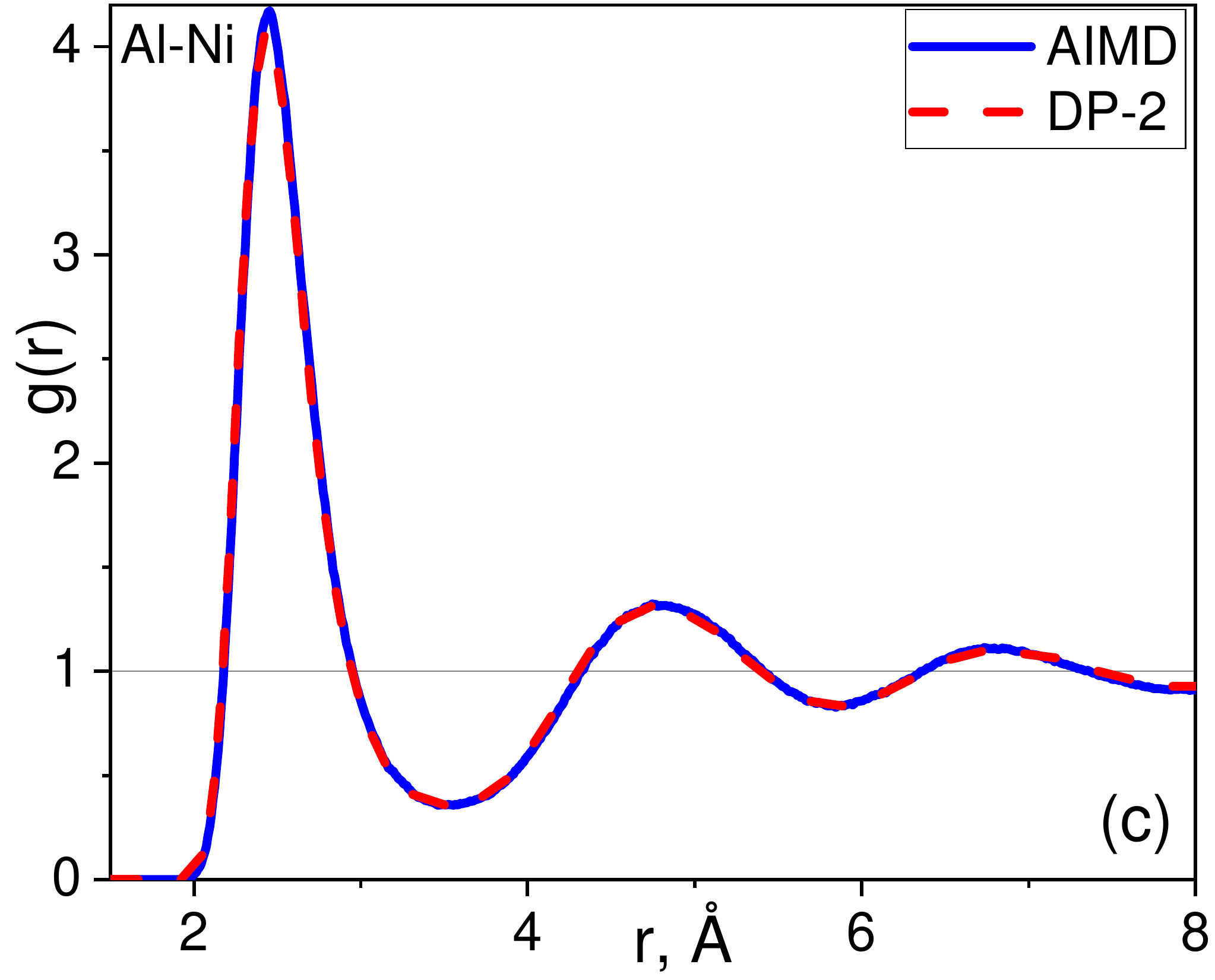}\\
 \includegraphics[width=0.25\textwidth]{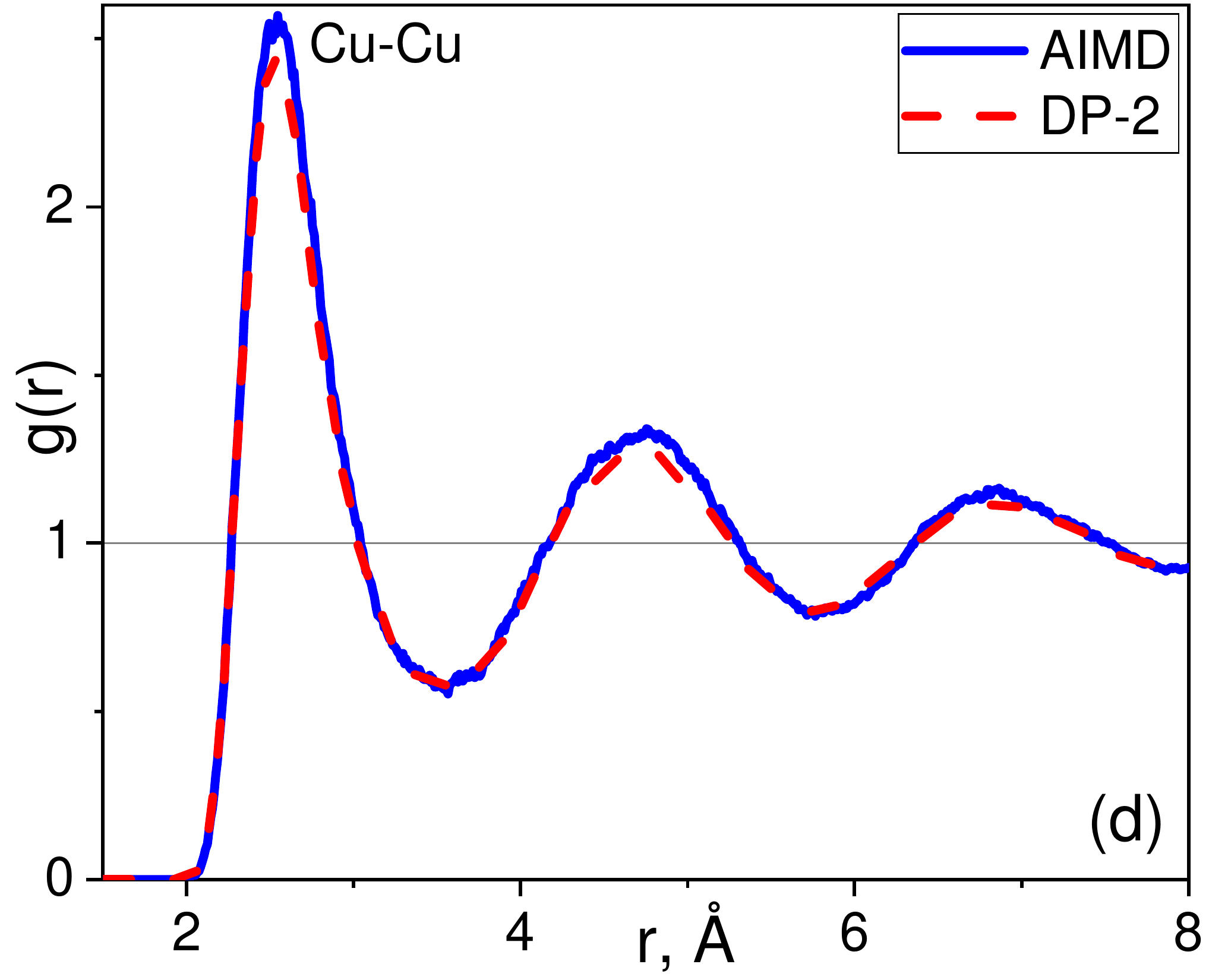}  \includegraphics[width=0.25\textwidth]{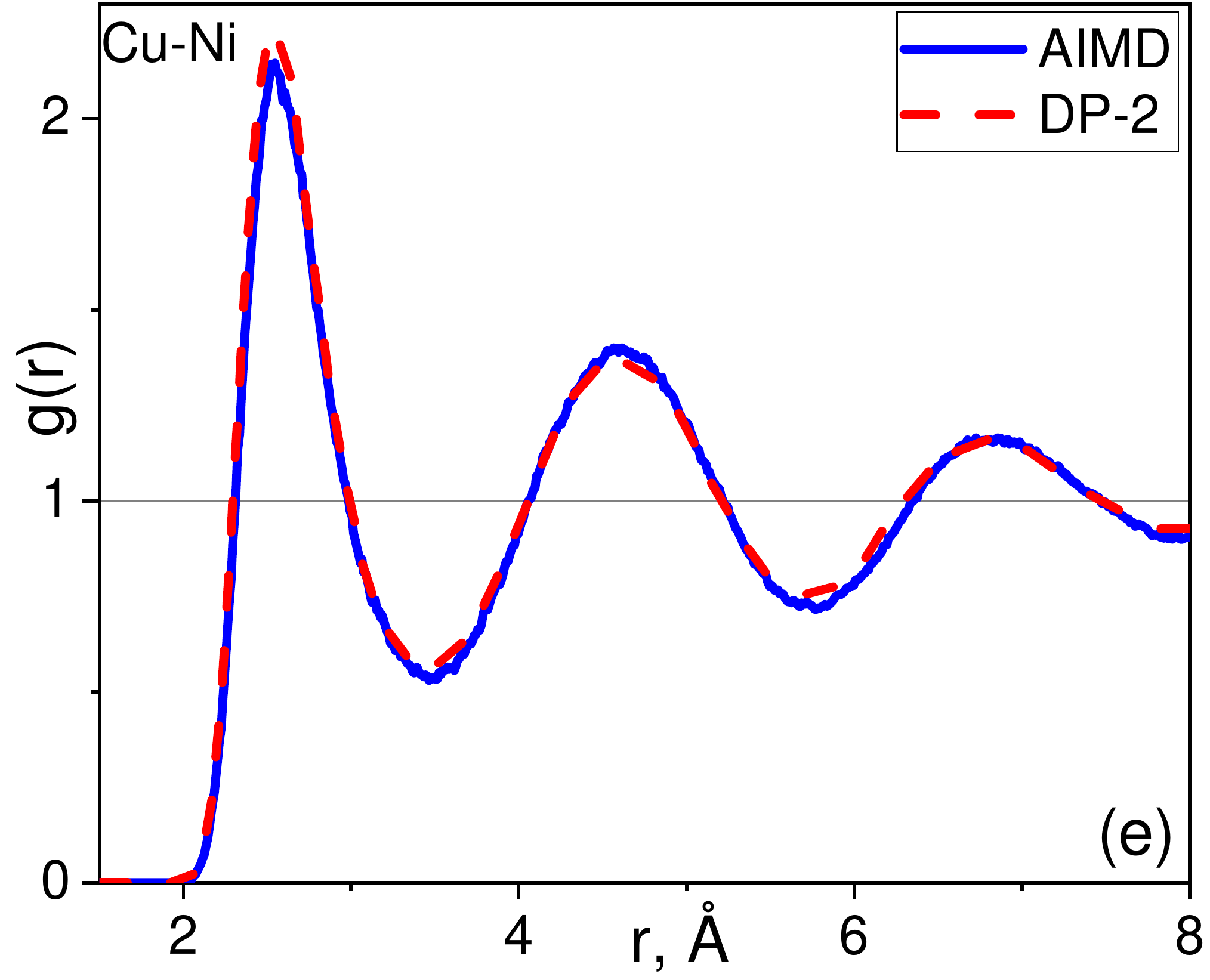}  \includegraphics[width=0.25\textwidth]{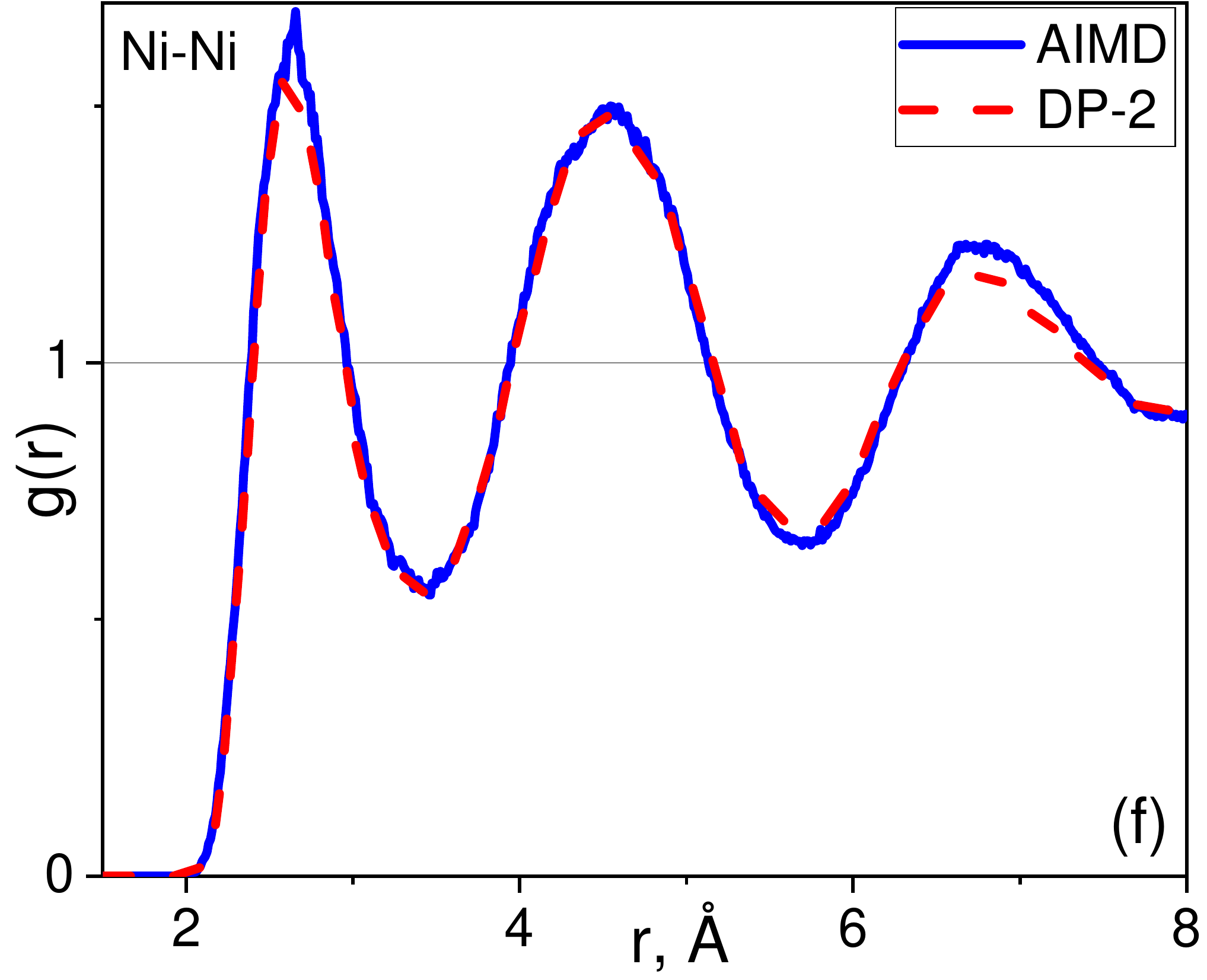}\\
 \includegraphics[width=0.25\textwidth]{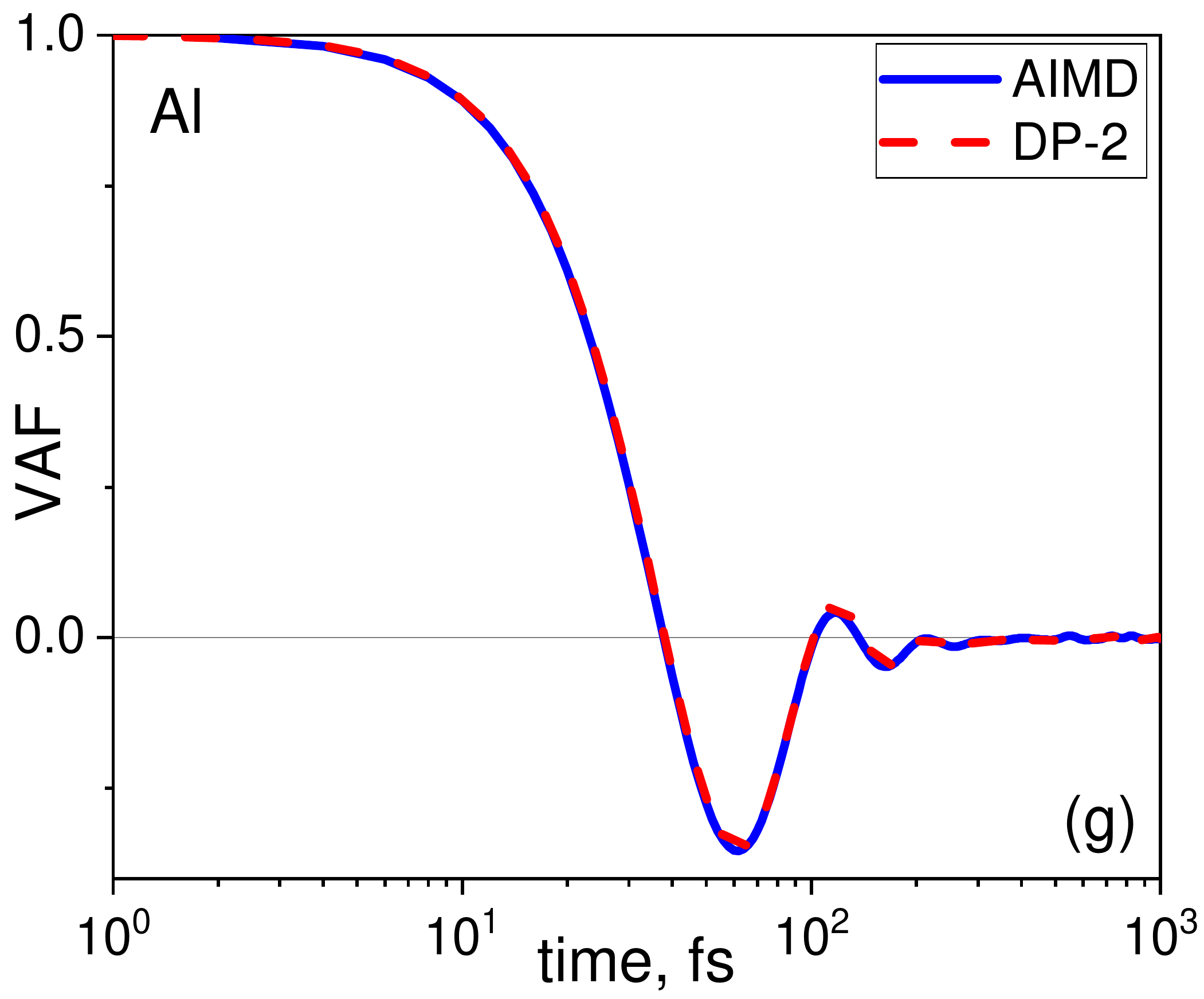}  \includegraphics[width=0.25\textwidth]{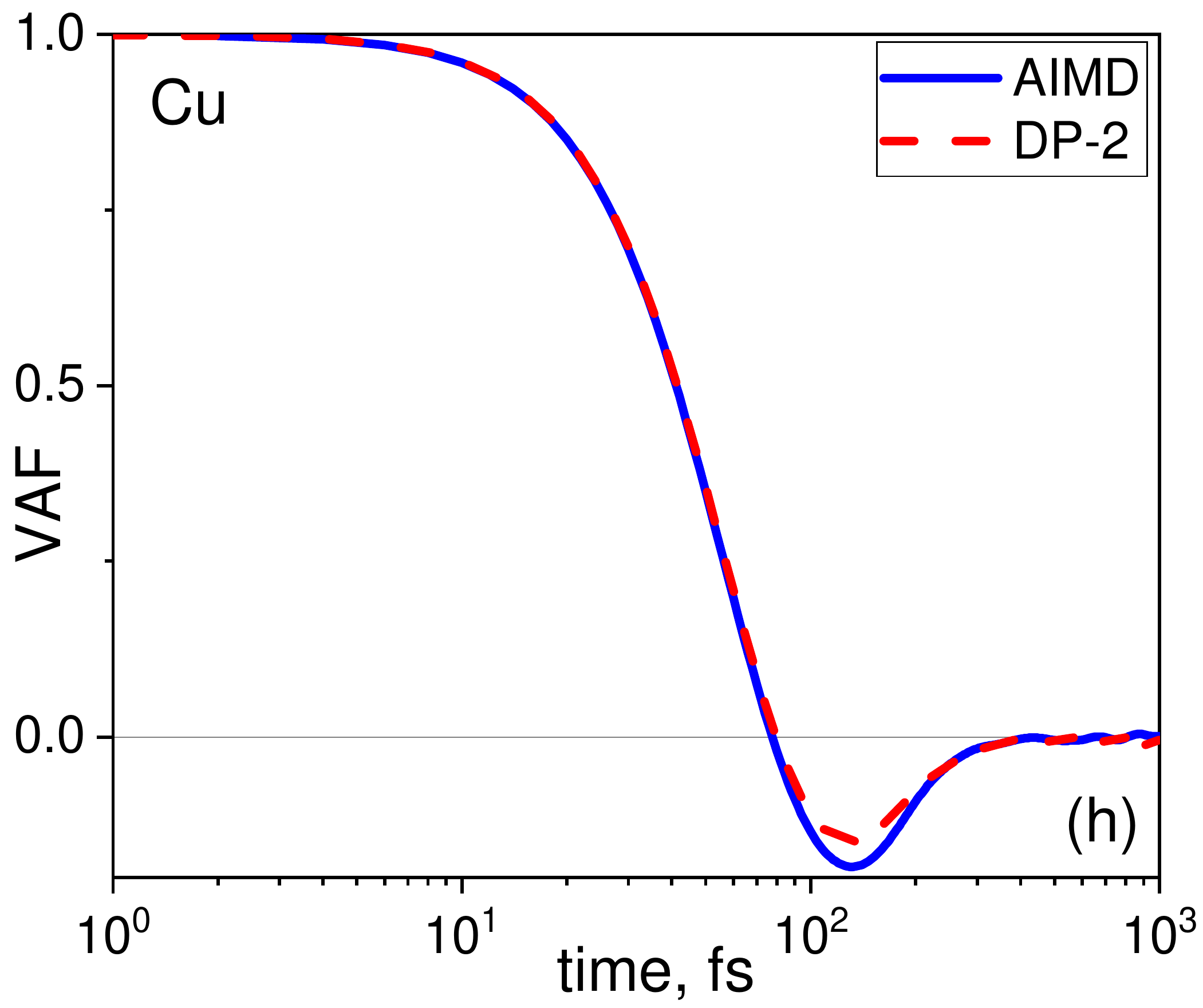}  \includegraphics[width=0.25\textwidth]{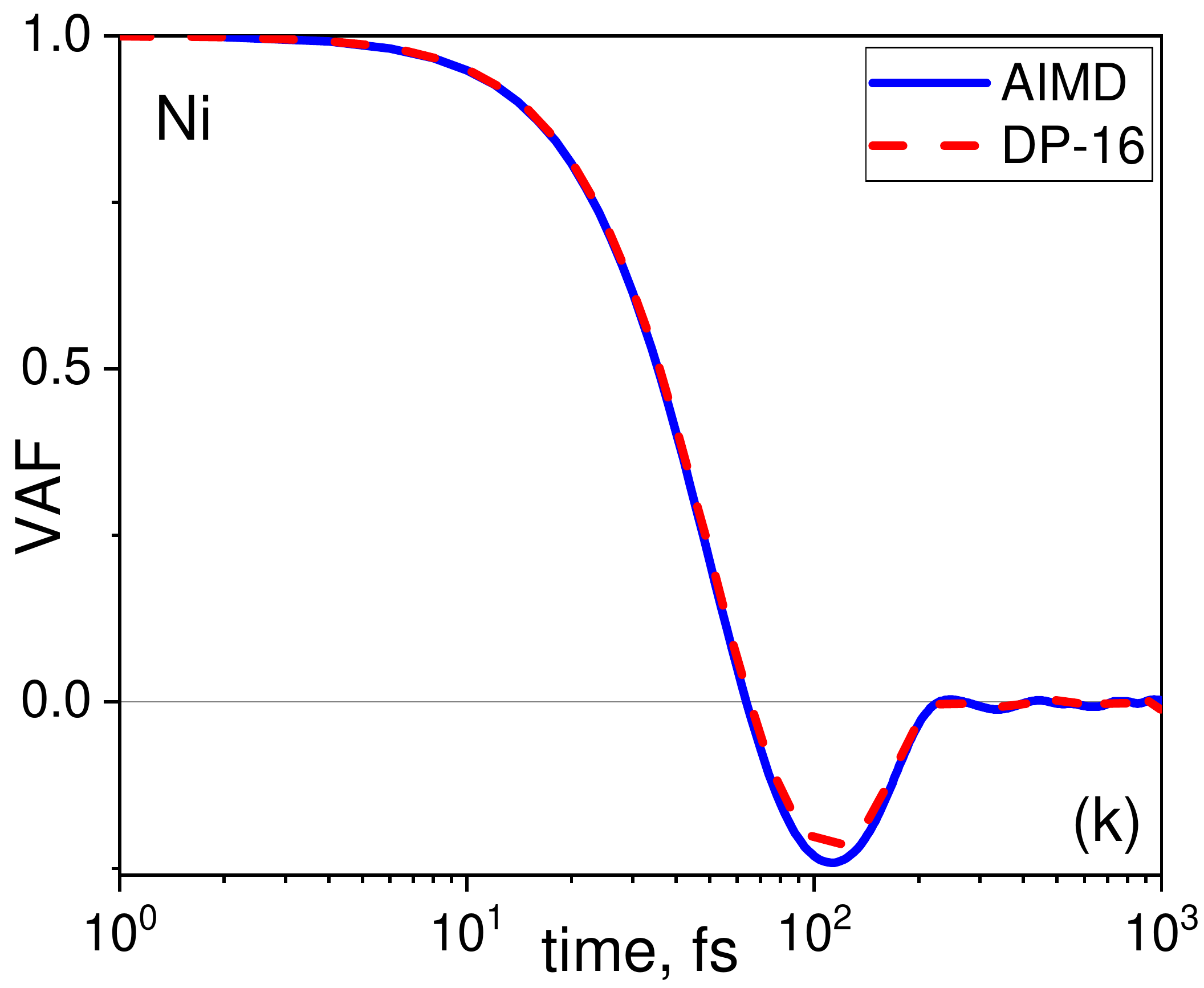}
  \caption{Partial radial distribution functions (a-f) and velocity autocorrelation functions (g-k) for ${\rm Al_{50}Cu_{25}Ni_{25}}$ melt at $T=1400$ K extracted from AIMD simulations (solid blue lines) as well as from DNNP simulations (red dashed lines).}
  \label{fig:RDF_VAF_test}
\end{figure*}

Thus, we see that developed DNNP reproduces well RDFs extracted from AIMD simulations in the whole concentration range. An important question arises how closely the potential describes experimental structural data. To address this point, we calculate RDFs for pure Al, Cu, and Ni, and compare the results with experimental data reported in Waseda's textbook~\cite{Waseda1980textbook}. Temperatures and densities in simulations were the same as in the experiments.  The results are shown in Fig.~\ref{fig:RDF_exp}. We see that DNNP provides an excellent description of experimental RDFs at temperatures, which are close to melting points; at elevated temperatures, the agreement becomes slightly worse but still good. This is an obvious result taking into account that training configurations were collected from AIMD simulations performed close to melting temperatures. For Ni, the agreement between experimental and simulated RDFs is slightly worse than for Al and Cu at all temperatures (see Fig.~\ref{fig:RDF_exp}d). At the same time, DNNP provides excellent agreement with AIMD data (compare black solid and dashed curves in Fig.~\ref{fig:RDF_exp}d). Thus, we argue the slightly worse description of experimental RDFs for Ni is rather an issue of agreement between DFT and experiment than the lack of DDNP accuracy. It may be caused by either underestimation of electronic correlations in Ni by DFT or by some problems with high-temperature XRD experiments.

Besides structural and dynamical properties discussed above, an important test for any interatomic potential is calculating thermodynamic properties. Here we address this issue for temperature dependencies of the density in liquid Al, Cu, and Ni. We perform NPT-ensemble simulations at different temperatures and compare the resulted equilibrium densities with experimental data reported in the literature~\cite{Waseda1980textbook,Brillo2003IntJThermPhys,Assael2006JPhysChemRefDat}. The results are presented in Fig.~\ref{fig:dens}. We see that, for liquid Al and Cu, DNNP describes density with experimental accuracy. For liquid Ni, the results are slightly worse, but the average deviation from experimental data is about 6\%. We suggest that the reasons for lower accuracy in describing density and structure of liquid Ni are the same and they are not related to the DNNP accuracy.

In Fig.~\ref{fig:dens}b we also illustrate the importance of virials in the training procedure. We see that DP-1, which has been trained without using virials, demonstrates much less accuracy in describing density than DP-2 for which the virials were included (see Tab.~\ref{tab:model_parameters}). This result suggests that the values of virials (or stress tensors in other machine learning models) should be includes in the training procedure to build more balanced models.

 \subsection{Compositional transferability of DNNPs\label{sec:comp_trans}}
We have shown above that the developed DNNPs provide excellent accuracy in describing structural, dynamical, and thermodynamical properties of Al-Cu-Ni melts at compositions that were included in the training dataset. Now we focus on the compositional transferability of DNNPs. To address this issue we choose several compositions that are most distant from the training concentration points (see Fig.~\ref{fig:ternary_plot}). First, we calculate DFT vs DDNP dependencies of energies, forces, and virials for that compositions and draw them together with those for training compositions (Fig.~\ref{fig:correlation}). We see that testing compositions also demonstrate linear correlations between DFT and DNNP data and their RMSE are similar to those for training compositions.  Then, we calculate partial RDFs and VAFs using DDNP-based simulations and compare them with the results of AIMD simulations. In Fig.~\ref{fig:RDF_VAF_test}, we show the results for ${\rm Al_{50}Cu_{25}Ni_{25}}$ alloy (other systems demonstrate similar results). We see that the agreement between DNNP and AIMD is as well as for the compositions included in the training dataset.

 Thus, we see that DNNP reveals good compositional transferability. This is an interesting and not obvious conclusion. Indeed, it is well known that all machine learning regressors (including neural networks) are not very good at extrapolation. However, there is no intuitive way to predict how far from the training compositions we can robustly use DDNPs. The results obtained show that we can use DDNPs at concentrations that are at least 20 at.\%. away from the training compositions. The question arises if this effect is either the result of extrapolation or it is due to compositional fluctuations. Indeed, although each configuration in the training dataset has a fixed composition, the configurations used in the training procedure belong to the spheres of a cutoff radius, which is usually less than the simulation box length. Actual concentrations of the components in these spheres can differ from those in the whole simulation box. However, direct estimations reveal that such compositional fluctuations do not exceed 4-5 at\%. Thus, the range of robust extrapolation is much more than the range of compositional fluctuations in the cutoff sphere.

\section{Conclusions\label{sec:discuss}}

In this paper, we have systemically explored the development of deep neural network potentials (DNNPs) for multicomponent metallic melts. Considering Al-Cu-Ni alloys as a convenient example and DeePMD-kit as a powerful tool, we focus on the following issues: (i) the search for configurations of neural networks and learning schemes that are optimal for simulating the metallic melts with respect to accuracy and computational efficiency; (ii) predictability of the developed DNNP; (iii) compositional transferability of the DNNP that is the possibility to describe compositions that are distinct from those included in the training dataset.

Considering different sets of DNNP hyperparameters as well as different learning schemes, we build fourteen DNNPs, which substantially differ in their accuracy and performance (see Tab.~\ref{tab:model_parameters}).

 All DNNPs were built on top of relatively short ($\sim$20~ps) AIMD trajectories produced for local order investigations. We did not apply special tricks for producing datasets like data decimation and also avoided an active learning strategy using the DPGEN tool. Our results show that for metallic melts even straightforward AIMD simulations produce reliable datasets for developing DNNPs.

Analyzing the developed DNNPs, we select an optimal model (DP-2) and examine carefully how it describes structural, dynamical, and thermodynamic properties of Al-Cu-Ni melts in comparison with the results of AIMD simulations and experimental data. We show that the DNNP reproduces very well DFT data on energy, forces, and virials (Fig.~\ref{fig:correlation}), as well as partial radial distribution functions (Fig.~\ref{fig:rdfs_aimd}) and velocity autocorrelation functions (Fig.~\ref{fig:vaf_test}) extracted from AIMD simulations. Experimental data on pure alloy components are also reproduced with good accuracy with the exception of Ni for which the agreement is slightly worse but still satisfactory (Figs.\ref{fig:RDF_exp}, \ref{fig:dens}); probably, this is due to a discrepancy between \textit{ab initio} and experimental data, not the drawback of the DNNP.

Surprisingly, we found that DNNP describes well the properties of the alloys whose concentrations are rather distinct from the compositions included in the training dataset. In other words, DDNP demonstrates good compositional transferability. The important consequence of this result is that DDNP developed at rather crude concentration mesh allows describing properties of the ternary alloy in the whole composition domain (see also Ref.~\cite{Jiang2021ChinPhysB} devoted to similar problem). These interesting results raise many questions. Is the compositional transferability a universal feature of DNNPs?  Does compositional transferability take place also for describing properties of solid phases including intermetallic compounds? Does the behavior of ternary alloys studied here extend to systems with more components? Obviously, a lot of further studies are needed in this direction.

Our findings open up prospects for simulating multicomponent metallic melts via DNNP. Creating interatomic potentials for high-performance and high-accuracy simulations of multiconmonets alloys in wide compositional ranges will give us effective tools for design and prediction of new materials.

\section*{Acknowledgments}
This work was supported by the Russian Science Foundation (grant 18-12-00438). The numerical calculations are carried out using computing resources of the federal collective usage center 'Complex for Simulation and Data Processing for Mega-science Facilities' at NRC 'Kurchatov Institute' (ckp.nrcki.ru/), supercomputers at Joint Supercomputer Center of Russian Academy of Sciences (www.jscc.ru), HybriLIT heterogeneous computing platform (LIT, JINR) \\
(http://hlit.jinr.ru) and 'Uran' supercomputer of IMM UB RAS (parallel.uran.ru).

\bibliographystyle{model1-num-names}
\bibliography{bib_nnp}
\end{document}